\documentclass[prd,nofootinbib,preprint,superscriptaddress]{revtex4}

\usepackage{amsmath, amssymb, amsthm, graphicx, epsfig, fancyhdr,epsfig, slashed}

\usepackage{tikzsymbols}
\usepackage{natbib}
\usepackage{float}
\usepackage{xcolor}

\usepackage{enumitem}
\usepackage{amsmath}

\usepackage{slashed}

\usepackage{subcaption}
\usepackage{braket}

\usepackage{hhline}
\usepackage{multirow}

\definecolor{lime}{HTML}{A6CE39}
\DeclareRobustCommand{\orcidicon}{
	\begin{tikzpicture}
	\draw[lime, fill=lime] (0,0) 
	circle [radius=0.2] 
	node[white] {{\fontfamily{qag}\selectfont \tiny ID}};
	\draw[white, fill=white] (-0.0625,0.095) 
	circle [radius=0.007];
	\end{tikzpicture}
	\hspace{-2mm}
}

\foreach \x in {A, ..., Z}{\expandafter\xdef\csname orcid\x\endcsname{\noexpand\href{https://orcid.org/\csname orcidauthor\x\endcsname}
			{\noexpand\orcidicon}}
}

\newcommand{\be}{\begin{equation}}
\newcommand{\ee}{\end{equation}}
\newcommand{\bea}{\begin{eqnarray}}
\newcommand{\eea}{\end{eqnarray}}

\newcommand{\ba}{\begin{eqnarray}}
\newcommand{\ea}{\end{eqnarray}}
\newcommand{\bi}{\begin{itemize}}
\newcommand{\ei}{\end{itemize}}

\newcommand{\x}{\star}

\usepackage{lineno}
\usepackage[english]{babel}
\usepackage{amsmath,amssymb,amsfonts, bm,bbm, slashed, subdepth}
\usepackage{graphicx}
\usepackage{enumerate}
\usepackage{setspace}
\usepackage{booktabs, tabularx}
\usepackage{units}
\usepackage{color}
\usepackage{float}
\usepackage{multirow}
\usepackage[pscoord]{eso-pic}
\usepackage[normalem]{ulem}
\usepackage{import}
\usepackage{url}
\usepackage{tikz}
\usetikzlibrary{snakes}
\usepackage{cancel}
\allowdisplaybreaks[1]
\usepackage{scalerel}
\usepackage{soul}
\usepackage{hyperref}
\usepackage{braket}
\hypersetup{
	setpagesize=false,
	bookmarksnumbered=true,
	colorlinks=true,
	linkcolor=blue,
	citecolor=red,
	hypertexnames=true
}
\usepackage{spverbatim}
\usepackage{subcaption}
\usepackage{tcolorbox}

\DeclareUnicodeCharacter{2032}{\ensuremath{\prime}}
\def\ba{\begin{eqnarray}}
\def\ea{\end{eqnarray}}
\def\br{\begin{array}}
\def\er{\end{array}}
\def\be{\begin{equation}}
\def\ee{\end{equation}}
\begin{document}

\title{Probing Leptophobic Dark Sectors via Gravitational Wave Signatures}

\author{Taramati}
\email{taramati@iitbhilai.ac.in}
\affiliation{Department of Physics, Indian Institute of Technology Bhilai, Durg-491002, India}

\author{Lekhika Malhotra}
\email{lekhika.malhotra@iitb.ac.in}
\affiliation{Indian Institute of Technology Bombay, Mumbai 400076, Maharashtra, India}

\author{Zafri A. Borboruah}
\email{zafri123@iitb.ac.in}
\affiliation{Indian Institute of Technology Bombay, Mumbai 400076,Maharashtra,India}

\author{Sudhanwa Patra}
\email{sudhanwa@iitbhilai.ac.in}
\affiliation{Department of Physics, Indian Institute of Technology Bombay, Mumbai 400076, India}
\affiliation{Institute of Physics, Bhubaneswar, Sachivalaya Marg, Bhubaneswar 751005, India}

\begin{abstract}
    We study a minimally extended version of the Standard Model where baryon number is gauged with a $U(1)_B$ symmetry. This model can be made anomaly-free by adding a set of additional fermions. The lightest component of these fermions behaves as a viable dark matter candidate. We show that the spontaneous breaking of $U(1)_B$ symmetry can produce gravitational waves via bubble dynamics resulting from a first-order phase transition, which can be detected in future gravitational wave experiments like LISA and ET. Such gravitational wave signatures can be used as a probe to constrain the model in future observations and complement dark matter and collider searches. We perform a random numerical scan of the parameter space and derive the viable region consistent with theoretical bounds from running of the coupling constants, current experimental bounds from dark matter experiments such as LUX-ZEPLIN and XENONnT and sensitive to future gravitational wave experiments. We find that dark matter with mass of 8 -- 12 TeV is the most interesting to test in future gravitational wave as well as laboratory experiments. In the viable parameter space, the mass of the $Z'$ gauge boson associated with the $U(1)_B$ lies in the 16 -- 24 TeV range, and the mass of the scalar associated with the symmetry breaking lies around 1 -- 2.5 TeV scale. Recent results from LUX-ZEPLIN rules out mass scales below TeV in this model, while dark matter with mass larger than 12 TeV will not be sensitive to future GW experiments. Hence, the dark matter and mediator mass scales of interest are marginally accessible at current collider energies.
\end{abstract}

\maketitle
\tableofcontents
\flushbottom
\section{Introduction}

The Standard Model (SM) has been the most successful framework for understanding the building blocks of matter and their interactions, but it is far from the full story. It does not incorporate gravity in its description of particles~\cite{Einstein:1916vd}, cannot explain the observed matter-antimatter asymmetry~\cite{Planck:2018vyg}, offers no mechanism for neutrino masses and mixings~\cite{Super-Kamiokande:1998kpq, SNO:2002tuh, K2K:2002icj}, and does not account for the dark matter (DM) component of the Universe~\cite{Julian:1967zz, SDSS:2003eyi}. These limitations strongly suggest the existence of physics beyond the SM, leading to a rich landscape of theoretical models collectively known as Beyond Standard Model (BSM) physics. One particularly intriguing idea is to minimally extend the SM gauge group by promoting the accidental global symmetries of the SM, like baryon number (B) or lepton number (L), to gauge symmetries, leading to new dynamics and new particles~\cite{Pati:1973uk, Pati:1974yy, Mohapatra:1974gc, Mohapatra:1974hk, Georgi:1974sy, Georgi:1974yf, Georgi:1974my, DeRujula:1975qlm, Fritzsch:1974nn, Pais:1973mi, Duerr:2013dza, Patel:2022qvv, FileviezPerez:2010gw, FileviezPerez:2011pt, Dulaney:2010dj}. In these extended models, new gauge bosons and fermions emerge naturally, often paving the way for a viable dark matter candidate.

These BSM scenarios are difficult to test in laboratory experiments because they require very high energies. Gravitational waves (GWs) provide a complementary probe in this context. Since their first detection by LIGO in 2015~\cite{LIGOScientific:2016aoc}, GWs have become a valuable tool for studying the early Universe and searching for signatures of new physics. Recent results from the NANOGrav collaboration~\cite{NANOGrav:2023gor, NANOGrav:2023hvm, 2018PhRvL.120g1301H, NANOGRAV:2018hou, Aggarwal:2018mgp, Brazier:2019mmu, NANOGrav:2020bcs} confirms the presence of a stochastic gravitational wave background (SGWB) in the Universe. In particular, first-order phase transitions (FOPTs) associated with the spontaneous breaking of symmetries in BSM theories can generate SGWB that remain imprinted in spacetime~\cite{Caprini:2024gyk,Caprini:2024hue,Caprini:2009yp,Hindmarsh:2013xza,Huber:2008hg,Anderson:1991zb,NANOGrav:2023hvm}. These signals are largely unaffected by intervening matter or radiation, making them pristine messengers of high-scale dynamics~\cite{Schwaller:2015,Brietbach:2018,Hashino:2018zbe,Borah:2021,Dasgupta:2023zrh,Brdar:2025gyo,Brdar:2018num,Graf:2021xku,Karmakar:2023ixo}. In the context of dark matter, this connection is particularly compelling, since models that attempt to explain dark matter often introduce new symmetries, which may undergo FOPTs in the early Universe~\cite{Alanne:2020,Chen:2023,Abe:2023,Srivastava:2025oer,Bringmann:2023iuz,Hosseini:2023qwu,Borah:2021ftr,Borah:2021ocu,StatusUpdate:2023,Ghosh:2022fzp,Roy:2022gop,Hooper:2025fda,Chatterjee:2022pxf,Borah:2024emz,Abe:2023yte}. The resulting gravitational wave signatures serve not only as indirect evidence for the model but also as a means to constrain its parameter space. This makes gravitational wave astronomy a powerful tool that will complement dark matter searches and collider experiments. 

Dark matter continues to evade detection despite extensive searches. In particular, Weakly Interacting Massive Particles (WIMPs) in the GeV to TeV mass range have been a central focus of interest. However, the lack of positive signals in direct detection experiments like LUX~\cite{daSilva:2017swg}, XENON-100~\cite{XENON100:2012itz}, PANDAX-II~\cite{PandaX-II:2016vec, PandaX-II:2017hlx} and XENON-1T~\cite{XENON:2015gkh,XENON:2018voc}, as well as in indirect detection probes like PAMELA~\cite{PAMELA:2013vxg, PAMELA:2011bbe}, AMS-2~\cite{Corti:2014ria}, Fermi-LAT~\cite{Fermi-LAT:2009ihh} and IceCube~\cite{IceCube:2017rdn, IceCube:2018tkk}, have imposed stringent limits on the parameter space of many WIMP-based models. These constraints have sparked a shift toward more sophisticated constructions where dark matter can evade detection while still accounting for the observed relic abundance~\cite{Bhattacharya:2018fus, Bhattacharya:2017sml, Ghorbani:2014qpa, Dutta:2021uxd, Barman:2019oda, Konar:2020wvl, Konar:2020vuu, Sarazin:2021nwo, Ghosh:2021khk, Borah:2021rbx, Mishra:2021ilq, Borah:2022zim, Paul:2024prs,Borah:2023dhk,Borah:2025ubr}. It is now well established that purely singlet leptonic dark matter is largely excluded, as it tends to either overproduce relic density or yield spin-independent scattering cross-sections in conflict with experimental limits~\cite{Bhattacharya:2018fus,daSilva:2017swg, XENON:2015gkh, XENON:2018voc, Planck:2018vyg, WMAP:2003elm}. A promising alternative lies in mixed dark matter states, particularly those arising from the mixing of SM singlets and doublets. Such constructions allow for significant suppression of couplings to the Z boson, thereby weakening direct detection signals without disrupting thermal freeze-out~\cite{Taramati:2024kkn}.

With the above motivation, in this paper, we aim to use future GW experiments to constrain the parameter space of a simple leptophobic extension of the SM, incorporating a fermionic dark matter candidate. Leptophobic models naturally forbid proton decay at the renormalizable level and allow new physics near the TeV scale without invoking a high-scale desert~\cite{FileviezPerez:2018jmr,Pais:1973mi,Carone:1995pu,FileviezPerez:2011pt,FileviezPerez:2014lnj,Ma:2020quj, FileviezPerez:2010gw, Duerr:2013dza, FileviezPerez:2019jju, MurguiGalvez:2020mcc}. Such a framework not only ensures proton stability but also provides viable Majorana-type~\cite{FileviezPerez:2018jmr,Duerr:2013dza}, Dirac-type~\cite{Duerr:2014wra,ElHedri:2018cdm}, and vector-like~\cite{Duerr:2013lka,Ellis:2018xal} dark matter candidates and uncovers novel avenues for baryogenesis. In this work, we extend the SM gauge group with a $U(1)_B$ local symmetry. To make the model anomaly-free, we add additional fermions to the particle content. Two of these extra fermions: one SM singlet and one doublet, mix through the scalar field that breaks the $U(1)_B$ symmetry~\cite{Taramati:2024kkn}. The breaking of $U(1)_B$ symmetry not only gives masses to the dark sector but also leaves behind a $Z_2$ symmetry that stabilizes the dark matter candidate. Because this dark matter state is partially charged under the SM gauge group, it interacts with W and Z bosons, as well as the new $Z'$ from the broken $U(1)_B$, offering interesting phenomenology for colliders and dark matter searches. If the symmetry breaking is first-order in nature, it proceeds via nucleation of bubbles. The dynamics of these bubbles, how they grow, collide, and stir up the surrounding plasma, produce a stochastic background of GWs. Future GW detectors like LISA~\cite{Baker:2019nia}, DECIGO/U-DECIGO~\cite{Seto:2001qf, Yagi:2011wg}, BBO~\cite{Crowder:2005nr, Corbin:2005ny}, ET~\cite{Punturo:2010zz, Hild:2010id}, etc., will reach sensitivities to probe phase transitions at the TeV scale and beyond. We demonstrate that a viable parameter space exists where the phase transition associated with $U(1)_B$ breaking is sufficiently strong to leave a distinct signature in the GW spectrum, peaked at frequencies that these instruments are designed to detect, making GWs an indirect probe of the dark sector.

This work is structured as follows: in section~\ref{sec:model}, we present the model framework, mass generation, and dark matter. Next in section~\ref {sec:PT}, we evaluate the temperature-dependent effective potential for this model and define the phase transition properties and gravitational waves from the first-order phase transition. In section~\ref{sec:bounds}, we present the current bounds on this model from electroweak precision experiments, dark matter direct and indirect detection experiments, and collider experiments. In section~\ref{sec:numerical analysis}, we describe the methods we used for parameter scan of the model incorporating all the current bounds on the model while looking for a strong first-order phase transition with GW signatures detectable in future GW experiments. In this section, we show the results of the parameter scan and the complementarity of GW with laboratory experiments in probing this model. We conclude the paper in section~\ref {sec:conl}. Appendix~\ref{sec:relic} provides a brief description of the calculation of the relic density of the dark matter candidate. In Appendix \ref{app:beta}, we discuss the renormalization group evolution of the model parameters relevant for our numerical analysis.


\section{Model}
\label{sec:model}

We consider a minimal $U(1)_B$ extension of the SM where baryon number is gauged. The SM particle content along with their $U(1)_B$ charges are given in Table~\ref{table:1}. Due to gauged baryon number, the non-zero anomalies that arise are~\cite{Taramati:2024kkn, FileviezPerez:2010gw},
\begin{eqnarray}
&& \mathcal{A}[SU(2)^2_L \otimes U(1)_{B}] =\frac{3}{2}, ~~~  \mathcal{A}[U(1)^2_Y \otimes U(1)_{B}] =-\frac{3}{2}.
\end{eqnarray}

\begin{table}[htb!]
\begin{center}
\begin{tabular}{ccccc}
	\hline
SM Fermions	& $ SU(3)_C$ & $SU(2)_L$ & $U(1)_Y$ & $U(1)_{B}$	\\
	\hline
	\hline
$Q_{L} =  \begin{pmatrix}
             u_L \\ d_L
            \end{pmatrix}$	& $\textbf{3}$ & $\textbf{2}$ & $1/6$ & $1/3$	\\
 $u_R$	& $\textbf{3}$ & $\textbf{1}$ & $2/3$ & $1/3$\\
 $d_R$	& $\textbf{3}$ & $\textbf{1}$ & $-1/3$ & $1/3$\\
$\ell_L =  \begin{pmatrix}
             \nu_L \\ e_L
            \end{pmatrix}$	& $\textbf{1}$ & $\textbf{2}$ & $-1/2$ & $0$	\\
$e_R$	& $\textbf{1}$ & $\textbf{1}$ & $-1$ & $0$	\\
[1mm] \hline
Scalar\\
$H$ & \textbf{1} & $\mathbf{2}$ & $\phantom{+}1/2$  & $0$ \\
			\hline \hline
\end{tabular}
\end{center}
\vspace{-0.17in}
\caption{Transformations of the SM particles.}
\label{table:1}
\end{table}
To cancel these anomalies, we add a pair of $SU(2)_L$ doublet fermions $\Psi_{L}, \Psi_{R}$ and two pairs of singlet fermions $\chi_{L},\chi_{R}$ and $\xi_{L}^+,\xi_{R}^+$. To give these exotic fermions masses, we introduce a complex scalar singlet $S$ that breaks the $U(1)_B$ symmetry. The quantum charges of these fields are presented in Table~\ref{table:2}.
The non-zero anomalies are now given by,
\begin{eqnarray}
\mathcal{A}[SU(2)^2_L \otimes U(1)_B] &=&\frac{3}{2}+ \frac{1}{2}(B_1-B_2),\\
\mathcal{A}[U(1)^2_Y \otimes U(1)_B] &=&-\frac{3}{2}-\frac{1}{2}(B_1-B_2),
\label{eq:anom}
\end{eqnarray}
where $B_1,B_2$ are $U(1)_B$ quantum charges as shown in Table~\ref{table:2}. To ensure anomaly cancellation, the condition $B_1 - B_2 = -3$ must hold. Thus $B_1$ and $B_2$ can take various values~\cite{FileviezPerez:2019jju,Duerr:2013dza}. In this paper, we choose $B_1 = -1$ and $B_2 = 2$, which provide us with a Dirac-type fermionic dark matter as we will discuss later. 

\begin{table}[H]
\centering
\begin{tabular}{cccc}
\hline\hline
 Fermions  & $\text{SU(2)}_L$ & $\text{U(1)}_Y$  & $\text{U(1)}_{B}$\\
\hline 
\hline
$\Psi_L = \begin{pmatrix}
           \Psi^+_L \\ \Psi^0_L
          \end{pmatrix}
$ & $\mathbf{2}$ & $\phantom{+}1/2$  & $B_1$\\
$\Psi_R = \begin{pmatrix}
           \Psi^+_R \\ \Psi^0_R
          \end{pmatrix}$ & $\mathbf{2}$ & $1/2$  & $B_2$\\
$\xi_L^+$ & $\mathbf{1}$ & $1$ & $B_2$\\
$\xi_R^+$ & $\mathbf{1}$ & $1$ & $B_1$\\
$\chi_L$ & $\mathbf{1}$ & $0$ & $B_2$\\
$\chi_R$ & $\mathbf{1}$ & $0$ & $B_1$\\
[1mm] \hline
Scalar\\
$S$ & $\mathbf{1}$ & $0$ & $B_1-B_2$ \\
\hline\hline
\end{tabular}
\caption{Exotic fermions and scalars and their quantum charges.}
\label{table:2}
\end{table}
            
\subsection{Scalar sector}
\label{sec:scalarB}
The scalar potential of our model can be written as~\cite{MurguiGalvez:2020mcc, Pruna:2013bma, FileviezPerez:2019jju},
\begin{eqnarray}
\label{eq:VHS}
V(H,S)&=&
   -\mu^2_{H} H^\dagger H  + \lambda_{H} \big( H^\dagger H \big)^2
-\mu^2_{S} S^\dagger S  +\lambda_{S}\big(S^\dagger S\big)^2 + \lambda_{HS} \big(S^\dagger S\big) \big(H^\dagger H\big),
\end{eqnarray}
where $\lambda_{HS}$ is the portal coupling. The two scalars acquire vacuum expectation values (VEVs),
\begin{equation}
    \label{eq:vevs}
    \braket{S}=\frac{v_B}{\sqrt{2}},\quad\braket{H}=\begin{pmatrix}
    0\\
    v/\sqrt{2}
\end{pmatrix},
\end{equation}
where $v=246$~GeV and break the $U(1)_B$ and the electroweak symmetries. At the ground state, we can write the scalar fields as fluctuations around their respective VEVs as,
\begin{eqnarray}\label{eq:VEVs}
&&H=\frac{1}{\sqrt{2}}\begin{pmatrix}
    h^+ +i G^+\\
    v+\tilde{h}+i G_Z
\end{pmatrix}
,~~~~S=\frac{1}{\sqrt{2}}\left(v_B+\tilde{s}+i G_B\right),
\end{eqnarray}
where $\tilde{h},\tilde{s},G_Z,G_B$ are neutral while $h^+,G^+$ are charged real fields. Minimization of the potential with respect to the two VEVs eliminates the $\mu^2$ terms, giving the mass matrix for the neutral scalars as,
\begin{eqnarray}
 M^2_{HS}
=\left(\begin{matrix}
2\lambda_{H}\,v^2 &  \lambda_{HS}\, v v_B \\
 \lambda_{HS}\, v v_B &2\lambda_{S}\,v_B^2
 \end{matrix}\right)
\end{eqnarray}
This matrix has the eigenvalues,
\begin{eqnarray}
&&m^2_{h} = v^2 \lambda_H +v_B^2 \lambda_S  -\sqrt{ (v^2 \lambda_H -v_B^2 \lambda_S)^2+( \lambda_{HS} v v_B)^2}, \nonumber \\
&&m^2_{s} = v^2 \lambda_H +v_B^2 \lambda_S  +\sqrt{ (v^2 \lambda_H -v_B^2 \lambda_S)^2+( \lambda_{HS} v v_B)^2}.
\label{eq:exofer11}
\end{eqnarray}
Here $m_h\simeq125$~GeV is the mass of the SM Higgs particle. The scalar mixing angle $\theta$ is defined as,
\begin{equation}
\tan{(2\theta)}= \frac{\lambda_{HS} \,v v_B }{\lambda_{H}v^2-\lambda_{S}v_B^2}.
\end{equation}
The scalar masses can be written in terms of the mixing angle as,
\begin{eqnarray}
&&m^2_{h} = 2v^2\lambda_H \cos^2\theta + 2v_B^2\lambda_S \sin^2\theta + \lambda_{HS} v\, v_B \sin 2\theta, \nonumber \\
&&m^2_{s} =2v^2\lambda_H \sin^2\theta + 2v_B^2\lambda_S \cos^2\theta - \lambda_{HS} v\, v_B \sin 2\theta.
\label{eq:exofer12}
\end{eqnarray}
The Goldstone modes have vanishing masses at zero temperature,
\begin{equation}
\label{eq:GoldstoneMassesSSB}
m_{h^+}=m_{G^+}=m_{G_Z}=m_{G_B}=0.
\end{equation}
However, as we shall see in the next section, at high temperatures, they have field-dependent non-zero masses which contribute to the 1-loop corrections to the tree-level potential. These Goldstone modes are \textit{eaten up} by the gauge bosons of the theory to give masses to $W^\pm, Z$ bosons of the SM and the new $Z'$ boson associated with the $U(1)_B$ symmetry~\cite{Patra:2016ofq, Halzen:1984mc}.

The requirement for the potential to be bounded from below gives the following conditions~\cite{MurguiGalvez:2020mcc},
\begin{eqnarray}\label{eq:boundedness condition}
{ \lambda_H>0, \quad \lambda_S>0 \quad \text{and}  \quad\sqrt{\lambda_H\lambda_S}+\frac{1}{2}\lambda_{HS}>0.}
\label{eq:exofer9}
\end{eqnarray}
Additionally, the perturbativity of the couplings imposes:
\begin{eqnarray}\label{eq:perturbativity condition}
{ \lambda_H<4 \pi, \quad \lambda_S<4\pi \quad \text{and} \quad \lambda_{HS}<4\pi.}
\label{eq:exofer10}
\end{eqnarray}

\subsection{Gauge sector}
\label{sec:gauge}
\noindent The masses of the neutral gauge bosons in our theory are as follows~\cite{Halzen:1984mc},
\begin{eqnarray}
 && m_\gamma=0, \nonumber \\
 && m_Z =\frac{1}{2}v\sqrt{g^2+g'^2} =93 \pm 2 ~\text{GeV},  \nonumber \\
 &&  m_{Z'}=3 {g_B} {v_B},\label{eq:gaugemasses}
\end{eqnarray}
where $\gamma$ is the standard photon and $g,g'$ are the gauge couplings associated with the $SU(2)_L$ and the $U(1)_Y$ gauge groups. Refer to~\cite{Halzen:1984mc} for the numerical values of the masses of charged gauge bosons $W^{\pm}$, which have their normal SM values.

\subsection{Fermionic dark matter}
\label{sec:DMmixing}
After symmetry breaking, the exotic fermions in our model are odd under a remnant $Z_2$ symmetry while all the SM particles are even~\cite{Ghosh:2021khk}. The four exotic fermions give rise to two neutral and two charged physical states. If the lightest of these four states is neutral, it can behave like a dark matter candidate stabilized by the $Z_2$ symmetry. The Yukawa Lagrangian for these fermions can be written as,
\begin{eqnarray}
\label{eq:Lfermion}
\mathcal{L}_Y&=&-y_1 \overline{\Psi_{L}} \tilde{H}\xi_R^+-y_2 \overline{\Psi_{R}} \tilde{H}\xi_L^+-y_3 \overline{\Psi_{L}} {H}\chi_R -y_4 \overline{\Psi_{R}} {H}\chi_L\nonumber \\ &&-y_{\psi} \overline{\Psi_{L}} S \Psi_{R} -y_{\xi} \overline{\xi}_{L}^+ S^* \xi_{R}^+-y_{\chi} \overline{\chi_{L}} S^* \chi_{R},
\end{eqnarray}
where $\Tilde{H}=i\sigma_2 H^*$.
The above Lagrangian can be re-expressed in terms of mass matrices as,
\begin{eqnarray}
\mathcal{L}_{Y} &=&
\overline{\left(\begin{matrix} \xi_L^+ &\Psi_{L}^+
\end{matrix}\right)}
{\left(\begin{matrix}
M_{\xi} & M_2 \\
M_1 & M_{\Psi}
\end{matrix}\right)}
{\left(\begin{matrix}
  \xi_{R}^+ \\ \Psi_R^+
\end{matrix}\right)}  +
 \overline{\left(\begin{matrix}
\chi_L^0  & \Psi_{L}^0
\end{matrix}\right)}
{\left(\begin{matrix}
M_{\chi} & M_4 \\
M_3 & M_{\Psi}
\end{matrix}\right)}
{\left(\begin{matrix}
 \chi_R^0 \\  \Psi_{R}^0
\end{matrix}\right)} +h.c.,
\label{eq:nflag1}
\end{eqnarray}
where the different matrix elements are given as,
\begin{align}\label{eq:fermionmasses}
M_{1}=\frac{y_1 v}{\sqrt{2}} \quad M_{2}=&\frac{y_2 v}{\sqrt{2}}, \quad M_{3}=\frac{y_3 v}{\sqrt{2}},
\quad M_{4}=\frac{y_4 v}{\sqrt{2}}\nonumber\\
M_{\chi}=\frac{y_{\chi} v_B}{\sqrt{2}}, \quad &M_{\Psi}=\frac{y_{\psi} v_B}{\sqrt{2}},\quad M_{\xi}=\frac{y_{\xi} v_B}{\sqrt{2}} 
\end{align}
For simplicity, we consider $M_3=M_4$ and $M_1=M_2$. In the small mixing limit, the neutral components of the singlet and doublet fermions combine to give rise to the physical states $\Psi_1$ and $\Psi_2$ with masses given by,
\begin{equation}\label{eq:DMmasses}
    m_{\Psi_1} \simeq M_{\chi}+M_4\sin{2\theta_{\text{DM}}},\quad m_{\Psi_2} \simeq M_{\Psi}-M_4\sin{2\theta_{\text{DM}}}.
\end{equation}
where $\theta_{\rm DM}$ is the mixing angle defined by,
\begin{eqnarray}\label{ref:mixang}
\tan{2\theta_{\text{DM}}}= - \frac{2M_4}{M_\Psi-M_{\chi}} .
\end{eqnarray}
Similarly, the charged component of the doublet and the charged singlet mix to give rise to two physical states with mass eigenvalues,
\begin{equation}\label{eq:charged fermion masses}
    m_{\Psi_{1}^+} \simeq M_{\xi}+M_2\sin{2\theta_{p}},\quad m_{\Psi_{2}^+} \simeq M_{\Psi}-M_2\sin{2\theta_{p}},
\end{equation}
with the mixing angle given by,
\begin{equation}
\tan{2\theta_{p}}=-\frac{2M_2}{M_{\Psi}-M_{\xi}}
\label{eq:cflag2}
\end{equation}
For $\sin\theta_{\rm DM}=0$, both $\chi$ and $\Psi^0$ (neutral component of $\Psi$) behave as independent dark matter candidates. For convenience, we only consider a single dark matter candidate in this paper, i.e., $\sin\theta_{\rm DM}>0$. Smaller $\sin\theta_{\rm DM}$ signifies that the DM is mainly of singlet type, while a larger $\sin\theta_{\rm DM}$ implies that the primary contribution to the DM is from the doublet. We take $\sin\theta_p$ to be negligibly small for simplicity. Additionally, we consider $y_\xi\gg y_\psi \gg y_\chi$ such that $\Psi_1$ is always the DM candidate and the charged fermions can decay to DM. In addition, $\Psi_1^+$ is now the heaviest, required for collider phenomenology to be discussed in the next section. In the small $\sin\theta_{\rm DM}$ and $\sin\theta_p=0$ limit, $m_{\Psi_2^+}\simeq m_{\Psi_2}$.

\section{Phase transition}
\label{sec:PT}
In this section, we explore the parameter space of our model that gives rise to a strong first-order phase transition. Such phase transitions are associated with the nucleation of bubbles of true vacua that expand approximately at the speed of light and collide with each other, producing gravitational waves in the process. We study the GW signals that can be detected in future GW experiments such as LISA, BBO and ET, complementing DM and collider searches.
\subsection{The effective potential}\label{sec:effective potential}
The dynamics of the fields and the spontaneous breaking of $U(1)_B$ depend on the temperature-dependent effective potential, $V_{\rm eff}$ of the theory. From a phenomenological perspective, we are interested in the scenario where the $U(1)_B$ breaking scale is much higher than the electroweak breaking scale, $v_B\gg v$. In this limit, the tree-level potential can be written from Eq.~\eqref{eq:VHS},
\begin{equation}\label{eq:V0}
V_0(\varphi_S) =  -\frac{\mu^2_{S}}{2} \varphi_S^2  + \frac{\lambda_{S}}{4} \varphi_S^4
\end{equation}
where $\varphi_S=v_B+\tilde{s}$ is the real part of the complex field $S$ given in Eq.~\eqref{eq:VEVs}. The effective potential can be written as~\cite{Li:2020eun,Graf:2021xku,Bringmann:2023iuz}:
\begin{align}\label{eq:Veff}
V_{\mathrm{eff}}(\varphi_S, T)= V_0(\varphi_S) + V_{\mathrm{CW}}(\varphi_S)+V_{\mathrm{CT}}(\varphi_S)+V_{\mathrm{T}}(\varphi_S,T)+V_D(\varphi_S,T)
\end{align}
The different potential terms are explained below: $V_{\mathrm{CW}}(\varphi_S)$ is the 1-loop Coleman-Weinberg (CW) contribution, renormalized in the $\overline{\textrm{MS}}$ scheme in the Landau gauge, given by \cite{Coleman:1973jx, Bringmann:2023iuz}

\begin{align}
V_{\mathrm{CW}}(\varphi_S)=\frac{1}{64\pi^2}\sum_{i}n_i\left(-1\right)^{2s_i} m^4_i(\varphi_S)\left(  \log\left[\frac{m^2_i(\varphi_S)}{\Lambda^2}\right] - C_i \right).
\label{eq:VCW}
\end{align}
Here, $m_i(\varphi_S)$ are the field-dependent masses of all the bosonic and fermionic states (including Goldstones) in our model and the summation runs over $i$, denoting these states. The relevant gauge and fermion sector masses are given by equations~\eqref{eq:gaugemasses},~\eqref{eq:DMmasses} and \eqref{eq:charged fermion masses} by replacing $v_B$ with $\varphi_S$ and taking $v\approx 0$ while the scalar masses are given by,
\begin{equation}
m^2_{h} = -\mu_H^2+\frac{\lambda_{HS}\,\varphi_S^2}{2},\quad m^2_s=-\mu_S^2+3\lambda_S\,\varphi_S^2
\label{eq:scalarmasses}
\end{equation}
where $\mu^2_H,\mu^2_S$ are obtained by minimizing the tree-level potential with respect to the VEVs $v$ and $v_B$. The Goldstone bosons acquire masses,
\begin{equation}
    m^2_{h^+,G^+,G_Z}=-\mu^2_H+\frac{\lambda_{HS}\,\varphi_S^2}{2},\quad m^2_{G_B}=-\mu^2_S+\lambda_S\,\varphi_S^2
\end{equation}

In this calculation, $\Lambda$ represents the renormalization scale, which we set to be $v_B$. Although we are working in the Landau gauge, the mass eigenvalues of the Goldstones can be non-zero even for $T=0$, as these are evaluated at field configurations rather than the tree-level VEVs at zero temperature. The spin and the number of degrees of freedom (dof) of a particle $i$ is denoted by $s_i$ and $n_i$, respectively, e.g. $s_i=1$ for bosons and $1/2$ for fermions while $n_i=1$ for scalar states $h,s,h^+, G^+, G_Z, G_B$, $n_i=3$ for $Z'$, one of which is the longitudinal mode, $n_i=4$ for the Dirac fermions $\Psi_1,\Psi_2,\Psi_{1}^+,\Psi_{2}^+$. Moreover, in the $\overline{\textrm{MS}}$ scheme: $C_i=3/2$ for scalars and fermions and $5/6$ for vector bosons. Note that in the assumed limit $v\ll v_B$, our phase transition analysis and hence the GW spectrum is not sensitive to the mixing angles $\sin\theta_{\rm DM}$ and $\sin\theta_p$ from the previous section.

When we check the masses and mixings resulting from the effective potential, they deviate from their predicted tree-level values even at zero temperature. So we impose an appropriate counterterm to re-normalize the potential to preserve the tree-level values of the mixing and masses~\cite{Basler:2016obg, Quiros:1999jp, Graf:2021xku, Bringmann:2023iuz},
\begin{equation}
    V_{\text{CT}}(\varphi_S)=-\frac{\delta \mu_S^2 }{2} \varphi_S^2+\frac{\delta \lambda_S }{4} \varphi_S^4
    \label{eq:VCT}
\end{equation}
where the coefficients of the counterterms can be calculated by minimizing the zero-temperature effective potential with the counterterms and demanding,
\begin{eqnarray}
 \frac{\partial V_{\text{CT}}(\varphi_S)}{\partial \varphi_S}\bigg\rvert_{\varphi_S=v_B}= -\frac{\partial V_{\text{CW}}(\varphi_S)}{\partial \varphi_S}\bigg\rvert_{\varphi_S=v_B}  
\end{eqnarray}
\begin{eqnarray}
 \frac{\partial^2 V_{\text{CT}}(\varphi_S)}{\partial \varphi_S^2}\bigg\rvert_{\varphi_S=v_B}= -\frac{\partial^2 V_{\text{CW}}(\varphi_S)}{\partial \varphi_S^2}\bigg\rvert_{\varphi_S=v_B}  
\end{eqnarray}
Our choice of the renormalization condition makes sure that the zero-temperature VEV of the effective potential is the same as the VEV of the tree-level potential. Next we write the finite temperature one-loop quantum correction term~\cite{Dolan:1973qd, Quiros:1999jp, Bringmann:2023iuz, Graf:2021xku,Bittar:2025lcr},
\begin{align}
V_{\mathrm{T}}(\varphi_S, T)=\frac{T^4}{2\pi^2}\sum_{i}n_i J_i\left( \frac{m^2_i(\varphi_S)}{T^2} \right),
\label{eq:VT}
\end{align}
where the summation is again taken over all scalars, fermions and vector bosons (both transverse and longitudinal components). The thermal functions are defined by the following integrals~\cite{Dolan:1973qd, Parwani:1991gq,Arnold:1992rz, Quiros:1999jp,Carena:2008vj},
\begin{align}
\label{eq:JBF}
J_{B,F}(y^2)=\int^{\infty}_0 \mathrm{d}x\; x^2 \log \left( 1\mp e^{-\sqrt{x^2+y^2}} \right).
\end{align}
Here $\mp$ sign denotes bosons and fermions, respectively. At high temperatures, finite-temperature perturbation theory breaks down due to infrared divergences for bosonic modes~\cite{Athron:2023xlk} in the finite temperature potential when $T\gg m$. ``Daisy resummation'' is the typical method used to re-sum these states~\cite{Carrington:1991hz}. Here, we employ the Espinosa technique~\cite{Arnold:1992rz}, which gives the contribution,
\begin{align}
V_D(\varphi_S)=-\frac{T}{12\pi}\sum_{i} n_i \left[ \left(m^2_i(\varphi_S)+\Pi_i(T)\right)^{3/2}-\left(m^2_i(\varphi_S)\right)^{3/2} \right],
\label{eq:VD}
\end{align}
where $i$ now only runs over the scalar fields and the longitudinal component of the dark photon. The one-loop thermal masses $\Pi_{a}(T)$ in our model are given by,
\begin{align}
    \label{eq:thermal masses}
    \Pi_{\tilde{s}}(T) &=  \left( {\frac{\lambda}{3}+\frac{(y_{\chi}+y_{\xi}+y_{\psi})^2}{12}+\frac{g^2}{4}} \right)T^2\\
    \Pi_{G_B}(T) &=  \left( {\frac{\lambda}{3}+\frac{(y_{\chi}+y_{\xi}+y_{\psi})^2}{12}+\frac{g^2}{4}} \right)T^2\\
    \Pi_{Z'}(T)&=  \left( {\frac{1}{12}+\frac{2}{3}} \right)g^2T^2.
\end{align}
In order to avoid using the thermal mass of the DM fermion in our calculations, we only resum the Matsubara zero modes in the daisy potential, which are the ones creating the infrared divergences of Eq.~\ref{eq:VD}.

\begin{figure}[H]
\centering
\includegraphics[width=0.7\textwidth]{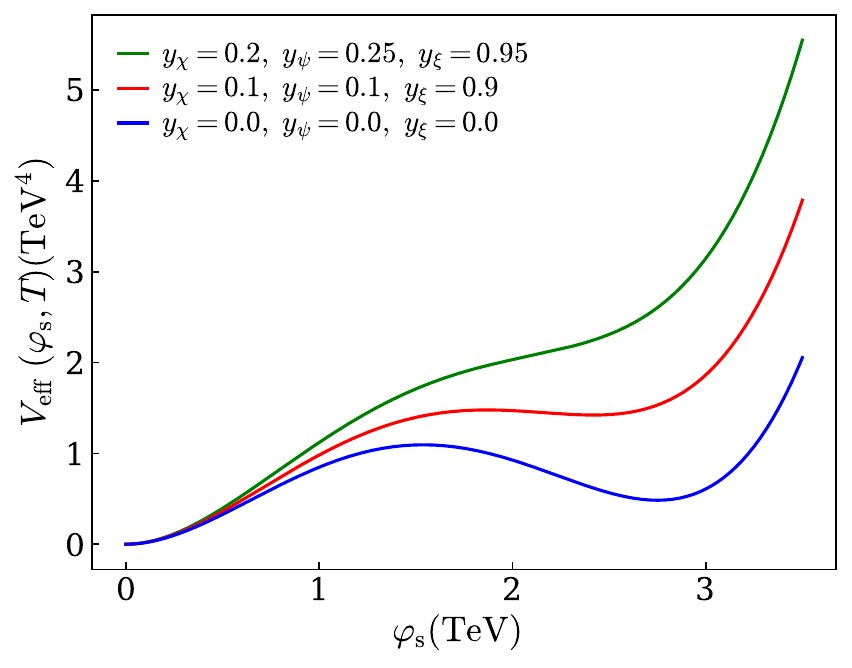}
\caption{\small Plot of $V_{\rm eff}(\varphi_S,T)$ given in Eq.~\eqref{eq:Veff} a temperature $T=2.27$ TeV, with quartic coupling $\lambda_S=10^{-3}$, $v_B=5$ TeV and $g_B=0.67$. Different colored lines represent different values of the Yukawa couplings $y_{\chi},y_\psi,y_\xi$.}
\label{fig:potential}
\end{figure}

In Fig.~\ref{fig:potential}, we show $V_{\rm eff}$ as a function of the field $\varphi_S$ at temperature $T=2.27$ TeV for a sample parameter point $\lambda_S=0.0035$ and $g_B=0.67$ while varying the Yukawa couplings $y_{\chi,\psi,\xi}$. The presence of exotic fermions tends to destabilize the Higgs vacua at $v_B$ for non-zero values of the Yukawa couplings due to their negative contribution in the Coleman-Weinberg corrections. To make sure that the effective potential is bounded from below, we check the condition that at zero temperature, $\varphi_S=v_B$ is the global minimum of the effective potential for each parameter choice we make, i.e. for large field values, $V_{\rm eff}(\varphi_S,0)>V_{\rm eff}(v_B,0)$.

\subsection{Phase transition properties}
The $U(1)_B$ symmetry is spontaneously broken at low temperatures at the global minima of the effective potential. This phase transition can be of first or second order, depending on the parameter values. First-order phase transition occurs through the nucleation of bubbles of the broken phase in the unbroken Universe. At higher temperatures, $V_{\rm eff}\propto T^2\varphi_S^2$; hence, the global minimum is at $\varphi_S=0$. As the temperature drops, another local minimum develops at $\varphi_S=v_B(T)$. At a critical temperature $T_c$, the two minima become degenerate. We are interested in identifying the subset of the parameter space for which there is a potential barrier between the two degenerate minima at $T_c$, which is necessary for a first-order phase transition. Below $T_c$, the temperature-dependent non-zero minima become the global minima while the Universe remains in the symmetric phase, which is now a metastable state. There is a finite probability that the field $s$ would tunnel to the global minima, bringing the Universe to the symmetry broken phase. The tunnelling rate is given by~\cite{Linde:1980tt, Coleman:1977py, Graf:2021xku}, 
\begin{equation}\label{eq:tunneling rate}
    \Gamma (T)\approx T^4\left (\frac {S_3}{2\pi T}  \right )^{3/2}e^{-\frac {S_3}{T}},
\end{equation}
where $S_3$ is the three-dimensional Euclidean action evaluated for the critical bubble corresponding to the bounce solution~\cite{Coleman:1973jx}. The bubble profile $s(x,T)$ can be calculated by solving the equation of motion,
\begin{equation}
    \label{eq:EOMbounce}
    \frac{d^2\varphi_S}{dx^2}+\frac{2}{x}\frac{d\varphi_S}{dx}=\frac{dV_{\rm eff}(\varphi_S,T)}{d\varphi_S}
\end{equation}
with boundary conditions $d\varphi_S/dx=0$ at $x=0$ and $\varphi_S\rightarrow0$ as $x\rightarrow\infty$ where $x$ is the 3D radial coordinate. $S_3$ is then defined as,
\begin{equation}\label{eq:S3}
    S_3=\int_{0}^{\infty}{\mathrm{d}x\mathrm{d}x^2\left [ \frac 12\left ( \frac {\mathrm{d}\varphi_S(x)}{\mathrm{d}x} \right )^2+V(\varphi_S,T) \right ]}.
\end{equation}

The transition is typically identified by the nucleation temperature $T_n<T_c$, which is the temperature at which the nucleation really takes place. It is defined as the temperature at which there is at least one nucleation per Hubble horizon,
\begin{equation} 
\label{eq:Tn}
    \int_{T_n}^{T_c}\frac {\mathrm{d}T}{T}\frac {\Gamma (T)}{\mathcal{H}(T)^4}=1.
\end{equation}
where $\mathcal{H}(T)$ is the Hubble parameter given as,
\begin{equation}
    \label{eq:hubble}
    \mathcal{H}(T)^2=\frac{\rho_{\rm rad}(T)+\rho_{\rm vac}(T)}{3M_{\rm pl}^2}=\frac{1}{3M_{\rm pl}^2}\left(\frac{\pi^2}{30}g_*T^4+\Delta V(T)\right)
\end{equation}
with $g_*$ is the relativistic degrees of freedom at temperature $T$ for our model and $M_{\rm pl}=2.435\times10^{18}$ GeV, the reduced Planck mass. The vacuum energy density is calculated from $\Delta V(T)=V_{\rm eff}(0,T)-V_{\rm eff}(v_B(T),T)$. In radiation-dominated epoch, the condition for nucleation temperature in Eq.~\eqref{eq:Tn} can be approximated by~\cite{Borboruah:2022eex},
\begin{equation}
    \label{eq:S3overT}
    \frac{S_3}{T_n}\sim -4\log \frac{1.66\sqrt{g_*}T_n}{M_{\rm pl}}\sim 100-140
\end{equation}
for $T_n\sim 10^6-10^2$ GeV. The nucleated bubbles are spherically symmetric and after nucleation, they expand and collide while interacting with the surrounding plasma. These collisions and plasma interactions are the sources of gravitational waves from FOPT. If the phase transition is of second order or not strongly first order, there is no tunnelling, hence no bubble formation. In that case, the field smoothly rolls down to the true vacuum at lower temperatures and no GWs are emitted.

In the case of FOPT, tunnelling does not imply that the whole Universe will transform into the symmetry-broken phase. Therefore, we need another temperature to characterize the phase transition, called the percolation temperature $T_p\lesssim T_n$, at which a significant portion of the Universe (approximately 70\%) is converted into the broken phase via merging of expanding bubbles. We calculate it by defining the probability of a point in space still being in the false vacuum at temperature $T$ to be $P(T)=e^{-I(T)}$, where $I(T)$ is the volume of true vacuum per unit comoving volume~\cite{Guth:1981uk},
\begin{equation}
    \label{eq:percI}
    I(T)=\frac{4\pi}{3}\int_{T}^{T_c}dT'\frac{\Gamma(T')}{T'^4\mathcal{H}(T')}\left(\int_T^{T'}\frac{d\tilde{T}}{\mathcal{H}(\tilde{T})}\right)^3
\end{equation}

From 3D percolation theory, the percolation temperature is calculated using the condition $I(T_p)=0.34$~\cite{Shante1971AnIT}. We find that $T_p \sim T_n$ for all the parameter points considered in our random scan, hence we calculate all the GW parameters at $T_n$.

To obtain the GW spectrum, we need to calculate two parameters, $\alpha$ and $\beta$. The parameter $\alpha$ measures the strength of the phase transition and corresponds to the vacuum energy released during the phase transition, normalized by the total radiation energy density~\cite{Espinosa:2010hh},
\begin{equation}\label{eq:alpha}
    \alpha=\frac{\rho_{\mathrm{vac}}}{\rho_{\mathrm{rad}}}=\frac{1}{\rho_{\mathrm{rad}}}\left [\frac T4 \frac {\mathrm{d}\Delta V}{\mathrm{d}T}-\Delta V \right ]_{T_n}
\end{equation}
Here, radiation energy density $\rho_{\mathrm{rad}}$ is given by $\rho_{\mathrm{rad}}=\frac {g_{\star}\pi ^2T^4}{30}$. The parameter $\beta$ signifies the inverse time duration of the phase transition~\cite{Kamionkowski:1993fg}, namely,
\begin{equation}
    \beta=\left (\mathcal{H}T\frac {\mathrm{d}(S_3/T)}{\mathrm{d}T}  \right )_{T_n}
\end{equation}

In our analysis, the effective potential is computed in the Landau gauge, which is a standard choice in studies of finite-temperature phase transitions as it simplifies the treatment of Goldstone modes and thermal corrections. We note, however, that the perturbative effective potential and quantities derived from it, such as the critical temperature $T_c$, nucleation temperature $T_n$, and the phase transition parameters $(\alpha,\beta/\mathcal{H})$, are in general gauge dependent. This issue has been extensively discussed in the literature, where it has been shown that while numerical values of these quantities may vary with gauge choice, the qualitative features of the phase transition and order-of-magnitude estimates of the resulting gravitational wave signals remain relatively robust. A fully gauge-invariant treatment would require the use of Nielsen identities or other gauge-invariant approaches, which is beyond the scope of the present work. We therefore interpret our results with this inherent theoretical uncertainty in mind. For detailed discussions, see Refs.~\cite{Dolan:1973qd,Jackiw:1974cv,Fukuda:1975di,Nielsen:1975fs,Patel:2011th,Garny:2012cg,Wainwright:2011qy,Wainwright:2012zn,Andreassen:2014eha,Andreassen:2014gha,Espinosa:2016nld}.

\subsection{Gravitational wave spectrum}
First-order phase transition in the early Universe could generate gravitational wave signals observable today. There are three different sources of gravitational waves produced in the first-order phase transitions: bubble wall collisions~\cite{Kosowsky:1991ua}, sound waves~\cite{Hindmarsh:2013xza} and magnetohydrodynamic (MHD) turbulence~\cite{Caprini:2006jb} in the plasma, i.e., the total gravitational wave strength is given by the sum of these as:
\begin{equation}
    \Omega_{\mathrm{GW}}h^2\simeq\Omega_{\mathrm{sw}}h^2+\Omega_{\mathrm{turb}}h^2+\Omega_{\mathrm{coll}}h^2.
\end{equation}
In order to calculate the contribution from each of these sources, we need to distinguish between 3 different scenarios of bubble dynamics: runaway, non-runaway and runaway in vacuum~\cite{Espinosa:2010hh, Caprini:2015zlo}. Depending on the strength of interaction of the bubble wall with the plasma after nucleation, bubbles can either accelerate to attain a terminal velocity before collisions take place (non-runaway)~\cite{Caprini:2015zlo} or keep accelerating to reach approximately the speed of light (runaway)~\cite{Espinosa:2010hh, Caprini:2015zlo}. For super-strong phase transitions, plasma effects can be neglected and the walls reach the speed of light (runaway in vacuum). In the case of runaway bubbles, collisions play the dominant part in the GW spectrum because the bubble walls collide at nearly the speed of light. In a non-runaway scenario, sound waves in the plasma mostly contribute to the total GW spectrum since the wall velocities are small. The distinction between runaway and non-runaway scenarios is made by comparing the $\alpha$ parameter with the $\alpha_\infty$ parameter, defined as~\cite{Espinosa:2010hh,Caprini:2015zlo},
\begin{equation}
\alpha_\infty \simeq \frac{30}{24\pi^2}\,\frac{1}{g_* T_n^2}\sum_a c_a\,\Delta m_a^2,
\end{equation}
where the sum runs over particle species that are light in the symmetric phase and become heavy after the transition. Here $\Delta m_a^2$ denotes the corresponding mass-squared difference, and the coefficient $c_a$ equals $N_a$ for bosons and $N_a/2$ for fermions, with $N_a$ being the number of degrees of freedom. We take $g_* \simeq 106.75$ at $T \sim T_n$.

In our model, the parameter $\alpha$ is typically smaller than $\alpha_\infty$ within the viable FOPT region, corresponding to non-runaway bubble dynamics. However, near the supercooled limit, $\alpha$ exceeds $\alpha_\infty$, indicating a transition to the runaway regime where bubble collisions significantly contribute to the GW spectrum. Supercooled phase transitions produce GW amplitudes that are orders of magnitude larger than those in the non-supercooled case, leading to substantially higher signal-to-noise ratios (SNR). Consequently, from a detection standpoint in future GW observatories, the runaway scenario dominates, with bubble collisions providing the primary contribution to the GW signal.

\subsubsection{Bubble collisions}

Assuming the \emph{envelope approximation}~\cite{Kosowsky1993GravitationalCollisions}, the contribution from bubble collision is given by \cite{Caprini2018CosmologicalWaves,Huber2008GravitationalBubbles,Weir2016RevisitingCollisions},
\begin{equation}
    h^{2} \Omega_{\mathrm{env}}(f)=1.67 \times 10^{-5}\left(\frac{H_{*}}{\beta}\right)^{2}\left(\frac{\kappa_c \alpha}{1+\alpha}\right)^{2}\left(\frac{100}{g_{*}}\right)^{\frac{1}{3}}\left(\frac{0.11 v_{w}^{3}}{0.42+v_{w}^{2}}\right) \frac{3.8\left(f / f_{\mathrm{env}}\right)^{2.8}}{1+2.8\left(f / f_{\mathrm{env}}\right)^{3.8}}
\end{equation}
where $H_*=H(T_n)$ and the efficiency factor $\kappa_c$ is given by \cite{Kamionkowski1994GravitationalTransitions},
\begin{equation}
    k_{\mathrm{c}}=\frac{0.715 \alpha+\frac{4}{27} \sqrt{\frac{3 \alpha}{2}}}{1+0.715 \alpha}
\end{equation}
In the above equation, $v_w$ is the wall velocity, which we assume to be unity for calculations. The peak frequency is given by
\begin{equation}
    f_{\text {env }}=16.5 \times 10^{-6}\left(\frac{0.62}{1.8-0.1 v_w+v_w^2}\right)\left(\frac{\beta}{H_{*}}\right)\left(\frac{T_{n}}{100 \mathrm{GeV}}\right)\left(\frac{g_{*}}{100}\right)^{\frac{1}{6}}\text{ Hz}
\end{equation}

\subsubsection{Sound waves}
The movement of the domain walls~(DWs) through the plasma creates pressure waves in the plasma. The contribution of such sound waves to GW is given from numerical fit \cite{Caprini2018CosmologicalWaves, Hindmarsh2015NumericalTransition, Guo_2021},
\begin{equation}
    h^{2} \Omega_{\mathrm{sw}}(f)=2.65 \times 10^{-6}\left(\frac{H_{*}}{\beta}\right)\left(\frac{\kappa_{sw} \alpha}{1+\alpha}\right)^{2}\left(\frac{100}{g_{*}}\right)^{\frac{1}{3}} v_{w} \left(\frac{f}{f_{\mathrm{sw}}}\right)^{3}\left(\frac{7}{4+3\left(f / f_{\mathrm{sw}}\right)^{2}}\right)^{7 / 2}\Upsilon
\end{equation}
where $v_w$ is the wall velocity which we assume as \cite{Lewicki:2021pgr},

\begin{equation}\label{eq:vw}
  v_w=
\begin{cases}
    \sqrt{\frac{\Delta V(Tn)}{\alpha \rho_{\text{rad}}}},& \text{if } \sqrt{\frac{\Delta V(Tn)}{\alpha \rho_{\text{rad}}}}\leq v_J\\
    1,              & \text{otherwise}
\end{cases}
\end{equation}
where $v_J$ is the Jouguet velocity, $v_J=\frac{1}{\sqrt{3}}\frac{1+\sqrt{3\alpha^2+2\alpha}}{1+\alpha}$. The efficiency factor $\kappa_{sw}$ is given by \cite{Schmitz2020NewTransitions,Caprini2016ScienceTransitions},
\begin{equation}\label{eq:kappasw}
  \kappa_{\mathrm{sw}}=
\begin{cases}
    \frac{\alpha}{0.73+0.083 \sqrt{\alpha}+\alpha},& \text{if } v_w\geq v_w^\alpha\\
    \frac{6.9\alpha v_w^{6/5}}{1.36-0.037 \sqrt{\alpha}+\alpha},              & \text{if }v_w<v_w^\alpha
\end{cases}
\end{equation}
where,
\begin{equation}
 v_w^\alpha=\left[\frac{1.36-0.037 \sqrt{\alpha}+\alpha}{6.9(0.73+0.083 \sqrt{\alpha}+\alpha)}\right]^{5/6}
\end{equation}
Lastly the factor $\Upsilon=1-\frac{1}{\sqrt{1+2\tau_{sw}H_*}}$ is a suppression factor dependent on lifetime of sound waves $\tau_{sw}$ \cite{Guo_2021}. It can be parameterized by writing $\tau_{sw}\sim R_*/\overline{U}_f$, where $R_*=(8\pi)^{1/3}v_w/\beta$ and $\overline{U}_f=\sqrt{3\kappa_{\mathrm{sw}}\alpha/4}$ represents mean bubble distance and root-mean-squared fluid velocity \cite{Hindmarsh:2017gnf}.
The peak frequency is given by,
\begin{equation}
    f_{\mathrm{sw}}=\frac{1.9 \times 10^{-5}}{v_w} \left(\frac{\beta}{H_{*}}\right)\left(\frac{T_{*}}{100 \mathrm{GeV}}\right)\left(\frac{g_{*}}{100}\right)^{\frac{1}{6}} \mathrm{Hz}
\end{equation}

\subsubsection{Magnetohydrodynamic turbulence}
The contribution of turbulent motion in a fully ionized plasma to the gravitational wave (GW) spectrum is described by magnetohydrodynamic (MHD) modelling~\cite{Caprini2018CosmologicalWaves, Caprini2009TheTransition, Binetruy2012CosmologicalSources},
\begin{equation}
    h^{2} \Omega_{\mathrm{turb}}(f)=3.35 \times 10^{-4}\left(\frac{H_{*}}{\beta}\right)\left(\frac{\kappa_{\text {turb }} \alpha}{1+\alpha}\right)^{\frac{3}{2}}\left(\frac{100}{g_{*}}\right)^{1 / 3} v_{w} \frac{\left(f / f_{\text {turb }}\right)^{3}}{\left[1+\left(f / f_{\text {turb }}\right)\right]^{\frac{11}{3}}\left(1+8 \pi f / h_{*}\right)}
\end{equation}
where the turbulence efficiency factor is given by $\kappa_{\mathrm{turb}}=0.05\kappa_{\mathrm{turb}}$ \cite{Caprini2016ScienceTransitions}. The characteristic frequency scale is expressed as,
\begin{equation}
    h_{*}=16.5 \cdot 10^{-6}\left(\frac{T_{n}}{100 \mathrm{GeV}}\right)\left(\frac{g_{*}}{100}\right)^{1 / 6} \mathrm{~Hz}
\end{equation}
The peak frequency is given by,
\begin{equation}
    f_{\text {turb }}=\frac{2.7 \times 10^{-5}}{v_w} \left(\frac{\beta}{H_{*}}\right)\left(\frac{T_{*}}{100 \mathrm{GeV}}\right)\left(\frac{g_{*}}{100}\right)^{\frac{1}{6}}\mathrm{Hz}
\end{equation}
\begin{table}[H]
\centering
\begin{tabular*}{0.8\linewidth}{@{\extracolsep{\fill}}ccccc}
 \hline
 \hline
 & BP1 & BP2 & BP3 & BP4\\
 \cline{2-5}
$\lambda_{HS}$ & 0.01 & 0.01 & 0.01 & 0.01  \\
$v_B$ (TeV) & $3.84$ & $19.28$ & $165.24$ & 105.22 \\
$m_S$ (TeV) & 0.219& 1.1374 &10.3687 &  5.151 \\
$m_{Z'}$ (TeV) & 2.493& 12.344 & 117.52 & 39.43\\
$m_\chi$ (TeV) & 1.246& 6.1719 & 58.7 &  19.72\\
$m_\psi$ (TeV) & 1.50 & 7.304& 67.3 & 43.67\\
$m_\xi$ (TeV) & 1.275& 6.409 & 81.390 &  42.37 \\
 \hline
 \hline
\end{tabular*}
\caption{Benchmark points for strong FOPT\label{tab:BPFOPT}. The calculated GW parameters for these benchmark points are given in Table~\ref{tab:BPFOPTres} and the corresponding GW signals are plotted in Fig.~\ref{fig:BP_GW}.}
\end{table}
\begin{table}[H]
\centering
\begin{tabular*}{0.8\linewidth}{@{\extracolsep{\fill}}ccccc}
 \hline
 \hline
 & BP1 & BP2 & BP3 &BP4\\
 \cline{2-5}
$T_c/$GeV & 407.54 & 2183.25 & 19451.33 &  16948.20 \\
$T_n/$GeV & 72.38 & 699.48 & 7965.24 & 16552.45  \\
$v_c/T_c$ & 9.42 & 8.83 & 8.49 & 3.35  \\
$\alpha$ & 49.96 & 3.90& 1.55 & 0.00922  \\
$\alpha_\infty$ &  0.01 & 0.02 & 0.019 & 0.011304  \\
$\beta/H_*$ & 976.01 & 1196.44& 1251.48 & 17448.24  \\
$\sqrt{\frac{\Delta V(Tn)}{\alpha \rho_{\text{rad}}}}$ & 0.91 & 0.94 & 1.0 & 0.94 \\
$v_J$ & 0.99 & 0.0.98 & 0.95 & 0.650 \\
$v_w^\alpha$ & 0.199 & 0.214 & 0.233 & 0.331  \\

 \hline
 \hline
\end{tabular*}
\caption{Phase transition parameters calculated for the benchmark points for FOPT given in Table \ref{tab:BPFOPT}. The GW signals for these benchmark points are plotted in Fig.~\ref{fig:BP_GW}.\label{tab:BPFOPTres}}
\end{table}

\begin{figure}[H]
\centering
\includegraphics[width=\textwidth]{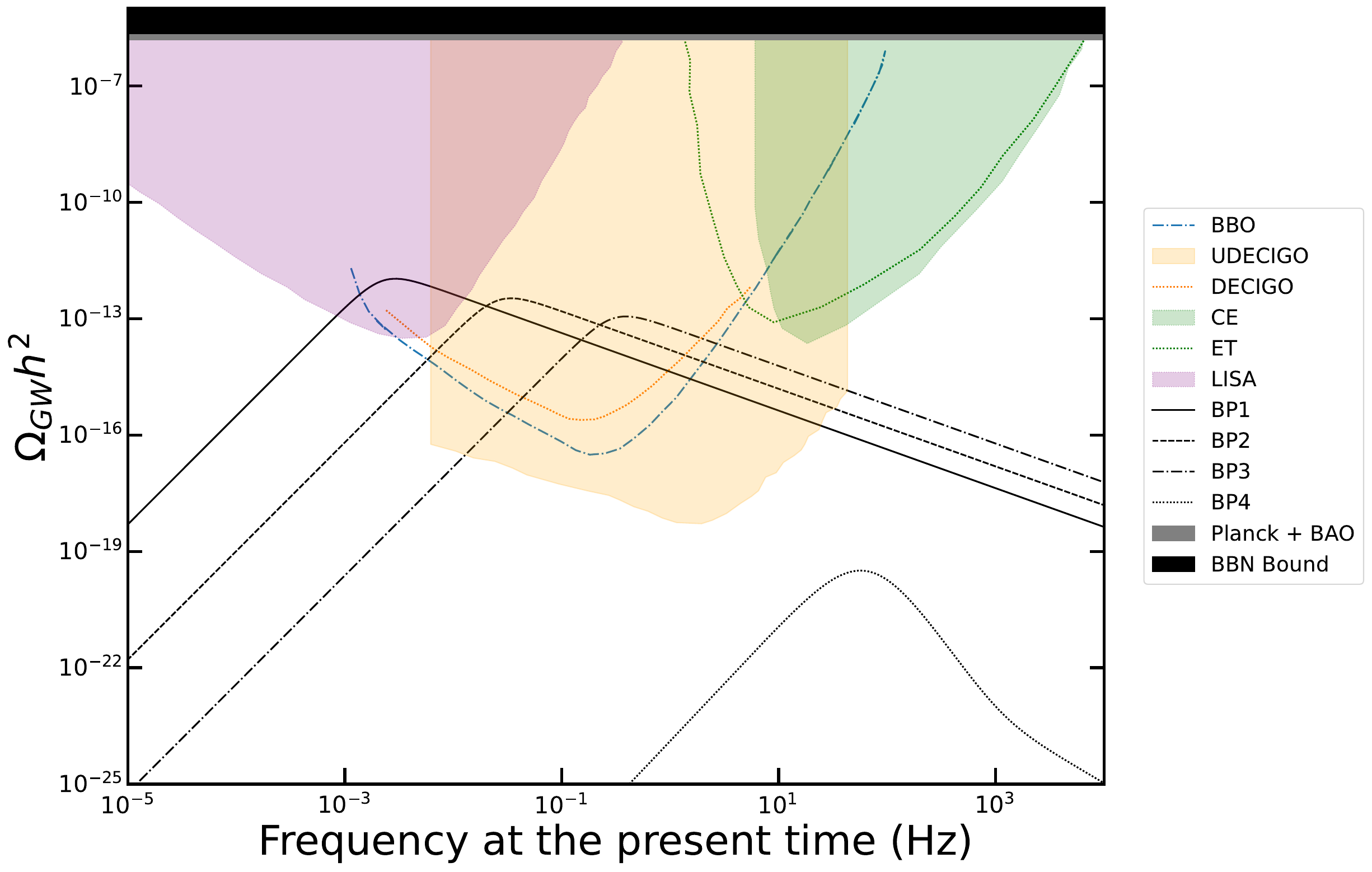}
\caption{GW spectra as seen today for benchmark points given in Table.~\ref{tab:BPFOPT} 
 and~\ref{tab:BPFOPTres} along with noise curves of future GW experiments.}
\label{fig:BP_GW}
\end{figure}

Fig.~\ref{fig:BP_GW} shows the GW spectrum from FOPT for the 4 benchmark points with $v_B\sim\mathcal{O}(1) -\mathcal{O}(100)$ TeV, as given in Table \ref{tab:BPFOPT}. The colored regions and lines show the sensitivity curves for various future GW experiments. We find that the peak GW frequency from this model in the considered parameter space is expected to be in the mHz-DeciHz range, detectable in experiments such as BBO~\cite{Yagi:2011wg}, LISA~\cite{amaroseoane2017laserinterferometerspaceantenna, Baker:2019nia}, DECIGO, U-DECIGO~\cite{Seto:2001qf,Yagi:2011wg}, ET~\cite{Punturo:2010zz,Hild:2010id}, CE~ \cite{LIGOScientific:2016wof,Reitze:2019iox} etc.

Gravitational waves contribute to the Universe's energy density as dark radiation, constrained by Big Bang Nucleosynthesis (BBN) and Cosmic Microwave Background (CMB) observations through $\Delta N_{\rm eff}$. The bound on the GW energy density is given by,

\begin{equation}
\int_{f_{\rm min}}^{\infty} \frac{df}{f} \Omega_{\rm GW}(f) h^2 \leq 5.6 \times 10^{-6} \Delta N_{\rm eff},
\end{equation}
A simplification of the above condition is $\Omega_{\rm GW} \leq 5.6 \times 10^{-6} \Delta N_{\rm eff}$. Figure~\ref{fig:BP_GW} shows BBN ($\Delta N_{\rm eff}^{\rm BBN} \simeq 0.4$ \cite{Cyburt:2015mya}) and Planck+BAO constraints ($\Delta N_{\rm eff}^{\rm Planck+BAO} \simeq 0.28$ \cite{Planck:2018vyg}). Future experiments like CMB-HD~\cite{CMB-HD:2022bsz}, CMB-Bharat~\cite{CMB-bharat}, CMB-S4~\cite{doi:10.1146/annurev-nucl-102014-021908} and NASA’s PICO project~\cite{Alvarez:2019rhd} aim for tighter limits $\Delta N_{\rm eff}^{\rm Proj.} = 0.014, 0.05$ and $0.06$ respectively.

The detection feasibility of the GW signal at an experiment is evaluated using the signal-to-noise ratio (SNR) defined by,

\begin{equation}\label{eq:SNR}
\text{SNR} = \sqrt{\tau \int_{f_\text{min}}^{f_\text{max}} \text{d}f \left(\frac{\Omega_\text{GW}(f) h^2}{\Omega_\text{exp}(f) h^2}\right)^2 },
\end{equation}
where $\tau = 4$ years and $f_{\rm min}, f_{\rm max}$ are the detector's frequency range. A signal is considered detected if $\text{SNR} \geq 10$.

We end this section by mentioning that in our setup, local cosmic strings will form when $U(1)_B$ is broken to $Z_2$, and contribute to the GW spectra~\cite{Vilenkin:2000jqa,Nielsen:1973cs,Kibble:1976sj,Vachaspati:1984gt,Gouttenoire:2019kij,Allen:1991bk,Auclair:2019wcv}. However, such GW signals are highly suppressed for symmetry breaking scales $v_B \sim 10^{4}\text{--}10^{5}\,\mathrm{GeV}$. The string tension is given by $\mu \simeq 2\pi v_B^2$, leading to a dimensionless parameter $G\mu \simeq 2\pi v_B^2/M_{\rm Pl}^2 \sim 10^{-30}\text{--}10^{-28}$. The resulting stochastic GW background from a scaling network of local cosmic strings is approximately flat with amplitude \cite{Vilenkin:2000jqa, Blanco-Pillado:2019vcs, Auclair:2019wcv,Ghoshal:2025iil}
\begin{equation}
\Omega_{\rm GW}^{\rm CS} h^2 \sim 10^{-2}\,\Omega_r h^2\,\sqrt{\alpha G\mu}
\sim 10^{-7}\sqrt{G\mu} \;\sim\; 10^{-22}\text{--}10^{-21},
\end{equation}
where $\Omega_r h^2 \simeq 4.2\times 10^{-5}$ and $\alpha \sim 0.1$. The characteristic frequency today is set by the formation temperature $T_{\rm form} \sim v_B$,
\begin{equation}
f_{\rm CS} \sim 10^{-3}\,{\rm Hz}\left(\frac{T_{\rm form}}{100\,\mathrm{GeV}}\right)
\sim 0.1\text{--}1\,\mathrm{Hz}.
\end{equation}

In contrast, a supercooled FOPT at the same scale, leading to runaway bubbles, produces a significantly stronger GW signal, as shown in Fig.~\ref{fig:BP_GW}. Here, BP1--BP3 correspond to runaway cases, while BP4 represents a non-runaway scenario with a GW amplitude comparable to that from cosmic strings. Since detectable SNRs in our analysis predominantly arise from runaway scenarios, we conclude that for $v_B \sim 10^{4}\text{--}10^{5}\,\mathrm{GeV}$,
\begin{equation}
\Omega_{\rm GW}^{\rm CS} \ll \Omega_{\rm GW}^{\rm FOPT},
\end{equation}
indicating that the GW contribution from local cosmic strings is negligible compared to that from the FOPT across all relevant frequencies.

\section{Current bounds}
\label{sec:bounds}

\subsection{Electroweak precision bounds}
\label{sec:EW}
Electroweak precision observables provide key insights into potential new physics beyond the SM~\cite{Peskin:1990zt, Peskin:1991sw, Kennedy:1990ib}. A powerful tool for capturing these effects is the $STU$ formalism\cite{Peskin:1991sw}, which expresses deviations in gauge boson self-energies through three oblique parameters: $S$ measures the impact of new physics on neutral current interactions at different energy scales, $T$ quantifies the relative shift in neutral versus charged current processes, linked to the $\rho$ parameter and $U$ captures additional effects on charged currents but is typically small. These parameters are defined as~\cite{Han:2000gp},

\begin{align}
\alpha S &= 4 s^2 c^2 \frac{\Pi_Z(m_Z^2) - \Pi_Z(0)}{m_Z^2}, \\
\alpha T &= \frac{\Pi_W(0)}{m_W^2} - \frac{\Pi_Z(0)}{m_Z^2}, \\
\alpha (S+U) &= 4 s^2 \frac{\Pi_W(m_W^2) - \Pi_W(0)}{m_W^2},
\end{align}
where $s$ and $c$ are the sine and cosine of the weak mixing angle and $\alpha$ is the fine structure constant, all evaluated at $m_Z$. The latest global fit from the Particle Data Group (2024)~\cite{ParticleDataGroup:2024cfk} constrains these parameters as follows:

\begin{equation}
S = -0.04 \pm 0.10, \quad T = 0.01 \pm 0.12, \quad U = -0.01 \pm 0.09.
\end{equation}
We evaluate $S$, $T$ and $U$ across various parameter points using \texttt{SPheno}~\cite{Porod:2003um}, identifying regions of parameter space that remain consistent with these bounds within $1\sigma$.

\subsection{DM direct detection}
In this section, we examine the constraints imposed by direct detection experiments on the viable parameter space of our DM model. These experiments are designed to probe potential interactions between DM particles and detector nuclei, typically resulting in measurable nuclear recoil signals. The relevant observable in this context is the spin-independent DM-nucleon scattering cross-section, $\sigma_{\rm SIDD}$, which we compute as a function of the DM mass $m_{\Psi_1}$.

In our model, the dark matter candidate interacts with nucleons primarily through processes mediated by $Z'$, 
$s$  and the SM $Z$  boson. However, the contributions from the $Z'$ and $s$ exchanges are suppressed due to the large masses appearing in their propagators, with the suppression becoming more pronounced as the $U(1)_B$ symmetry-breaking scale $v_B$ increases. Consequently, the dominant contribution arises from the $Z$-mediated interactions (for more details see \cite{Taramati:2024kkn}), where the scattering cross-section is governed by the mixing angle $\sin\theta_{\rm{DM}}$, which controls the singlet-doublet composition of the dark matter candidate. This makes $Z$-mediated processes particularly relevant for potential signatures in direct detection experiments.
The detailed expressions for the tree-level \texttt{SIDD} cross-section for a fermionic DM can be found in~\cite {FileviezPerez:2010gw,FileviezPerez:2011pt,FileviezPerez:2019jju,FileviezPerez:2016erl}. These interactions are sensitive to the mixing angle and mass spectrum of the exotic fermions, which also influence the thermal relic abundance. For our analysis, we employ the \texttt{micrOMEGAs}~\cite{Alguero:2023zol} framework, which incorporates nuclear form factors and astrophysical inputs to evaluate $\sigma_{\rm SIDD}$.

\begin{figure}[H]
\centering
\includegraphics[width=0.7 \textwidth]{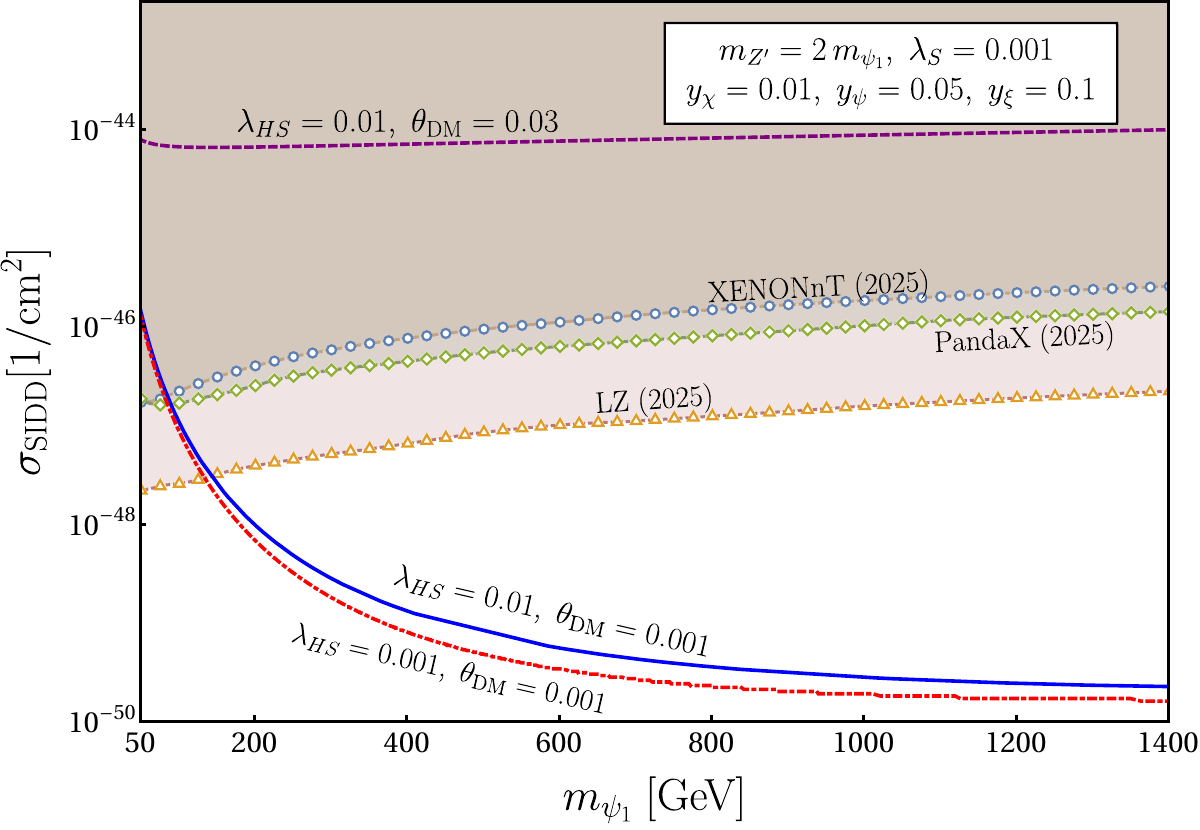}
\caption{Direct detection bounds from PandaX-4T (2025), LUX-ZEPLIN (2025) and future projection for XENONnT (2025) plotted along with $\sigma_{\rm SIDD}$ calculated for our DM candidate for various values of model parameters. Here we assume the resonant condition $m_{Z'}=2m_{\Psi_1}$.}
\label{fig:dd}
\end{figure}

In Fig.~\ref{fig:dd}, we plot $\sigma_{\rm SIDD}$ as a function of $m_{\Psi_1}$ for representative values of the input parameters, along with current direct detection bounds from PandaX-4T (2025)~\cite{PANDA-X:2024dlo}, LUX-ZEPLIN (LZ) (2025)~\cite{LZ:2024zvo} and XENONnT (2025)~\cite{XENON:2025vwd}. The most stringent bound comes from LZ (2025). The effects of $\lambda_{HS}$ and $\theta_{\rm DM}$ are shown. We see that for $\theta_{\rm DM}=0.001$, the DM candidate is predominantly singlet-like, suppressing its interactions with nucleons mediated by the SM $Z$ boson. This results in lower cross-section values, with most of the parameter points lying below the experimental bounds. In contrast, for $\theta_{\rm DM}=0.03$, the DM candidate acquires a larger doublet component, significantly enhancing its  $Z$-mediated interactions with nucleons. This leads to a notable increase in the cross-section for large $\theta_{\rm DM}$, causing DM candidates with masses below 1400 GeV to be ruled out by LZ (2025). For smaller values of $\theta_{\rm DM}$, LZ (2025) establishes a lower bound on the mass $m_{\Psi_1}$. In our random parameter scan, we implement this bound on our DM candidate. We also see that for a larger value of $\lambda_{HS}$, the cross-section is generally larger, which is expected since a larger coupling means a stronger mixing between DM and nucleons via scalar exchange, which in turn enhances the cross-section.

\subsection{DM indirect detection}
\label{sec:IDD}
Indirect detection of DM involves searching for SM particles produced in its annihilation or decay~\cite{Conrad:2014tla}. In the context of our model, the DM candidate is a Dirac-type fermion. Unlike Majorana DM scenarios, where Final State Radiation (FSR) is typically suppressed, Dirac-type DM produces a continuous gamma-ray spectrum rather than distinct monochromatic gamma-ray lines~\cite{Borboruah:2024lli}. As a result, searching for a sharp gamma-ray signature is challenging and the focus shifts to analyzing diffuse gamma-ray emissions arising from dark matter annihilation in astrophysical environments.

\begin{figure}[H]
\centering
    \begin{tikzpicture}[line width=0.5 pt, scale=0.85]
          \draw[solid] (-3.0,1.0)--(-1.5,0.0);
        \draw[solid] (-3.0,-1.0)--(-1.5,0.0);
         \draw[snake] (-1.5,0.0)--(0.8,0.0);
        \draw[solid] (0.8,0.0)--(2.5,1.0); \draw[snake] (0.8,0.0)--(2.5,1.0);
        \draw[snake] (1.59,0.5)--(2.5,0.0);
         \draw[solid] (0.8,0.0)--(2.5,-1.0); \draw[snake] (0.8,0.0)--(2.5,-1.0);
         \node at (-3.4,1.0) {${\rm DM}$};
         \node at (-3.4,-1.0) {$\overline{\rm DM}$};
         \node [above] at (0.0,0.05) {$Z'$};
        \node at (3.13,1.0) {$f/W^\pm$};
        \node at (2.7,0.0) {$\gamma$};
        \node at (3.2,-1.0) {$\overline{f}/W^{\mp}$};
         \draw[solid] (5.0,1.0)--(6.5,0.0);
        \draw[solid] (5.0,-1.0)--(6.5,0.0);
         \draw[snake] (6.5,0.0)--(8.5,0.0);
         \draw[solid] (8.5,0.0)--(10.4,1.0);\draw[snake] (8.5,0.0)--(10.4,1.0);
         \draw[solid] (8.5,0.0)--(10.4,-1.0); \draw[snake] (8.5,0.0)--(10.4,-1.0);
          \draw[snake] (9.5,-0.5)--(10.4,0.0);

         \node at (4.6,1.0) {${\rm DM}$};
         \node at (4.6,-1.0) {$\overline{\rm DM}$};
         \node [above] at (7.4,0.05) {$Z'$};
         \node at (10.99,1.0) {$f/W^\pm$};
        \node at (10.99,-1.0) {$\overline{f}/W^\mp$};
        \node at (10.5,0.0) {$\gamma$};
     \end{tikzpicture}
\caption{Tree-level annihilation channels of dark matter leading to the production of diffuse gamma-ray signals. In these processes, the final state includes fermion pairs $f\bar{f}$, where $f$ may be a Standard Model or an exotic fermion predicted by the model. }
\label{fig:FeynIdd}
\end{figure}
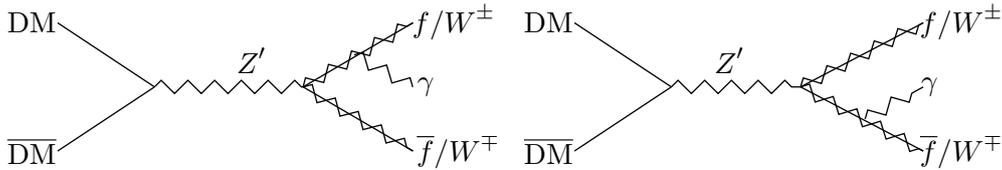

In our model, the dominant annihilation channels proceed via the exchange of the $Z'$ boson, leading to final states that predominantly include quark-antiquark pairs (e.g. $b \bar{b}$, $t \bar{t}$ ) and gauge bosons ($W^+ {W^-}$), or leptons ($\tau \bar{\tau}$). Fig.~\ref{fig:feyidd1} and \ref{fig:FeynIdd} shows example processes. These annihilation products contribute to the diffuse gamma-ray flux observed by current experiments such as \texttt{Fermi-LAT}~\cite{Fermi-LAT:2015kyq,Fermi-LAT:2015bhf,Foster:2022nva,Fermi-LAT:2011vow}, \texttt{H.E.S.S.}~\cite{HESS:2016mib,HESS:2018cbt} and next-generation observatory \texttt{CTA}~\cite{CTAConsortium:2017dvg,CTAO:2024wvb}. These experiments provide stringent constraints on the thermally averaged annihilation cross-section $\braket{\sigma v}_{\rm ann}$ of the DM. We use \texttt{micrOmegas} to calculate the annihilation cross section as a function of DM mass.

\begin{figure}[H]
\centering
    \begin{tikzpicture}[line width=0.5 pt, scale=0.85]

          \draw[solid] (-1.5,1.2)--(-0.5,0.0);
        \draw[solid] (-1.5,-1.2)--(-0.5,0.0);
         \draw[snake] (-0.5,0.0)--(1.3,0.0);  \draw[dashed] (-0.5,0.0)--(1.3,0.0);
        \draw[solid] (1.3,0.0)--(2.6,0.7);
        \draw[solid] (1.3,0.0)--(2.6,-0.7);
        \draw[solid] (2.6,0.7)--(2.6,-0.7);
        \draw[snake] (2.6,0.7)--(3.2,1.2);
         \draw[snake] (2.6,-0.7)--(3.2,-1.2);
         \node at (-1.7,1.2) {${\rm DM}$};
         \node at (-1.7,-1.2) {$\overline{\rm DM}$};
         \node [above] at (0.5,0.05) {$Z/Z', h_i$};

        \node at (2.1,0.0) {$f$};
        \node at (3.3,1.2) {$\gamma$};
        \node at (3.5,-1.2) {$\gamma, Z$};
         \draw[solid] (5.5,1.2)--(6.5,0.0);
        \draw[solid] (5.5,-1.2)--(6.5,0.0);
         \draw[snake] (6.5,0.0)--(8.3,0.0);  

         \draw[solid] (8.3,0.0)--(9.5,0.7);
        \draw[solid] (8.3,0.0)--(9.5,-0.7);
        \draw[solid] (9.5,0.7)--(9.5,-0.7);

         \draw[snake] (9.5,0.7)--(10.4,1.2);
         \draw[dashed] (9.5,-0.7)--(10.4,-1.2);
         \node at (5.3,1.2) {${\rm DM}$};
         \node at (5.5,-1.2) {$\overline{\rm DM}$};
         \node [above] at (7.6,0.05) {$Z'$};
          \node at (9.0,0.0) {$f$};
         \node at (10.5,1.2) {$\gamma$};
        \node at (10.5,-1.2) {$h$};
     \end{tikzpicture}
\caption{Loop-induced annihilation of dark matter particles into monochromatic photons through the processes ${\rm DM}\hspace{0.02cm}\overline{\rm DM} \to \gamma \gamma$, $\gamma Z$ and $\gamma h$ in a generic BSM framework featuring an exotic gauge boson mediator~($Z'$). The internal fermion $f$ represents either a Standard Model fermion or an exotic fermionic state predicted by the model.
}
\label{fig:feyidd1}
\end{figure}
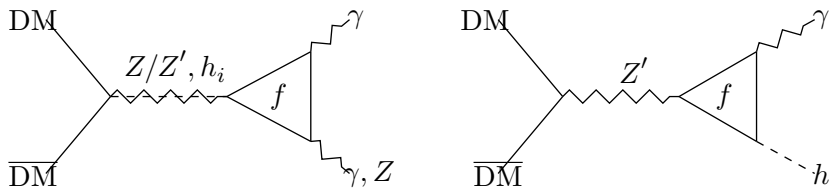

\begin{figure}[H]
\centering
\includegraphics[width=0.48 \textwidth]{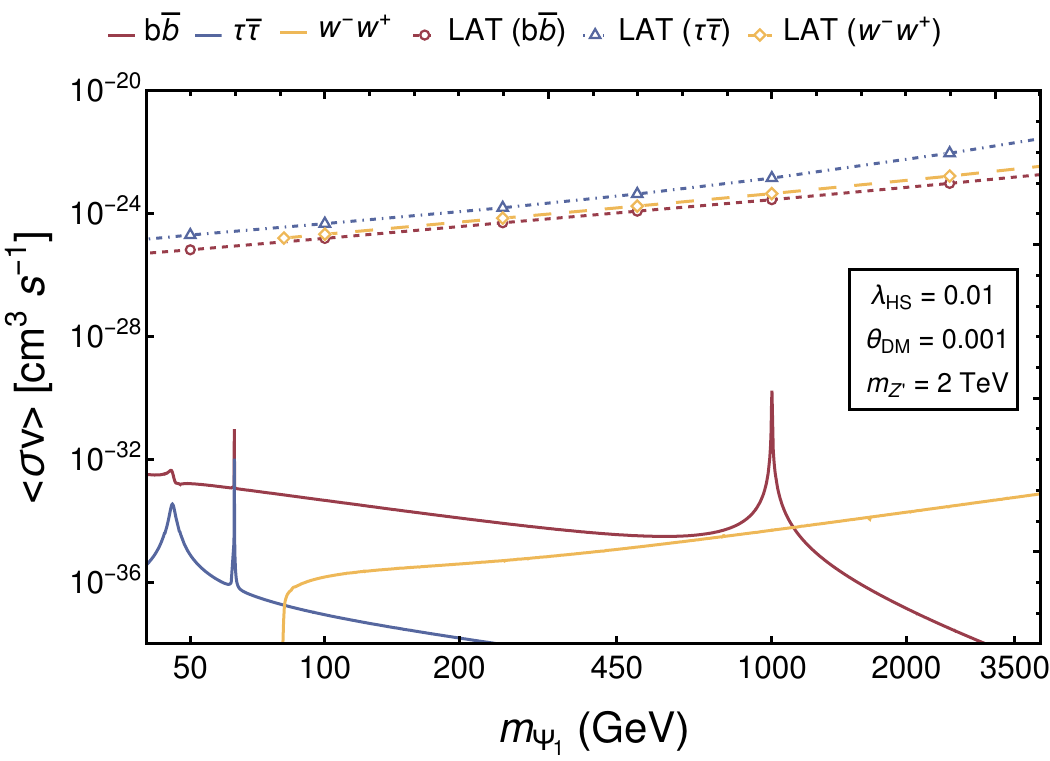}
\includegraphics[width=0.48 \textwidth]{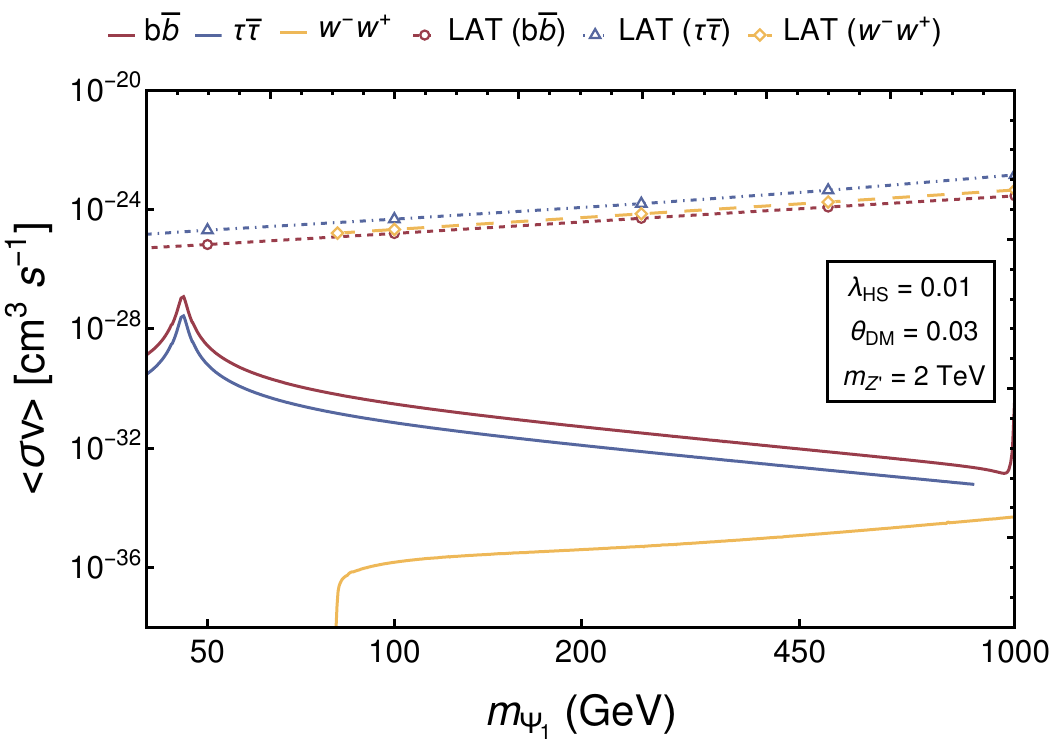}
\includegraphics[width=0.48 \textwidth]{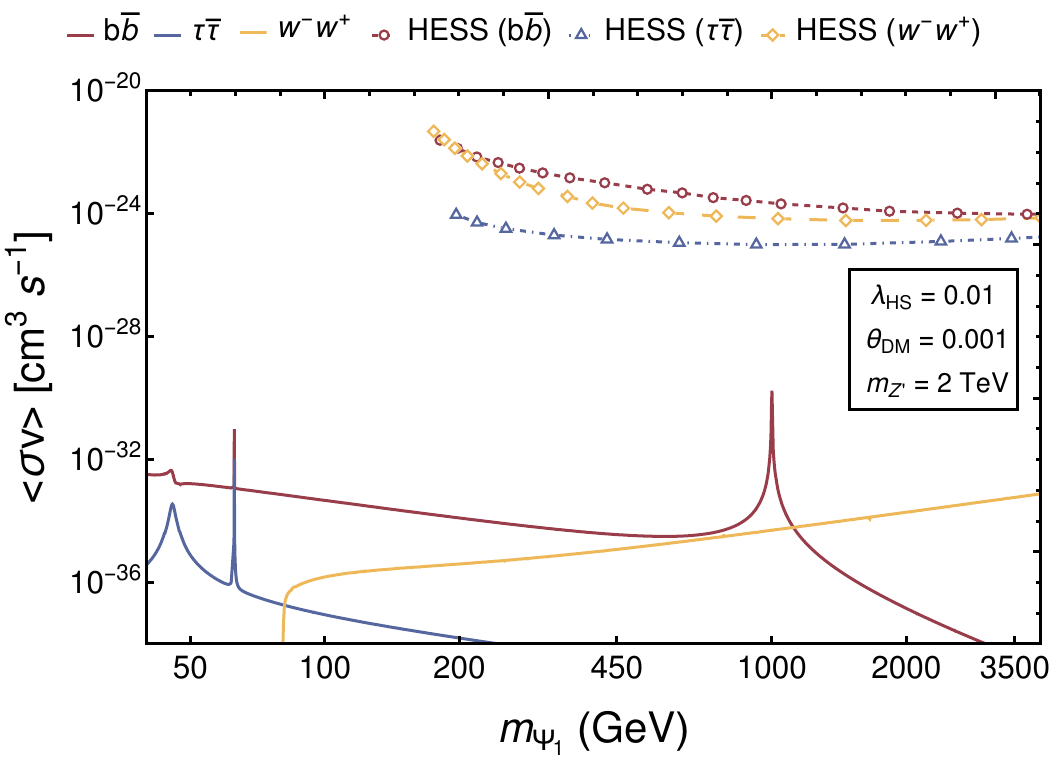}
\includegraphics[width=0.48 \textwidth]{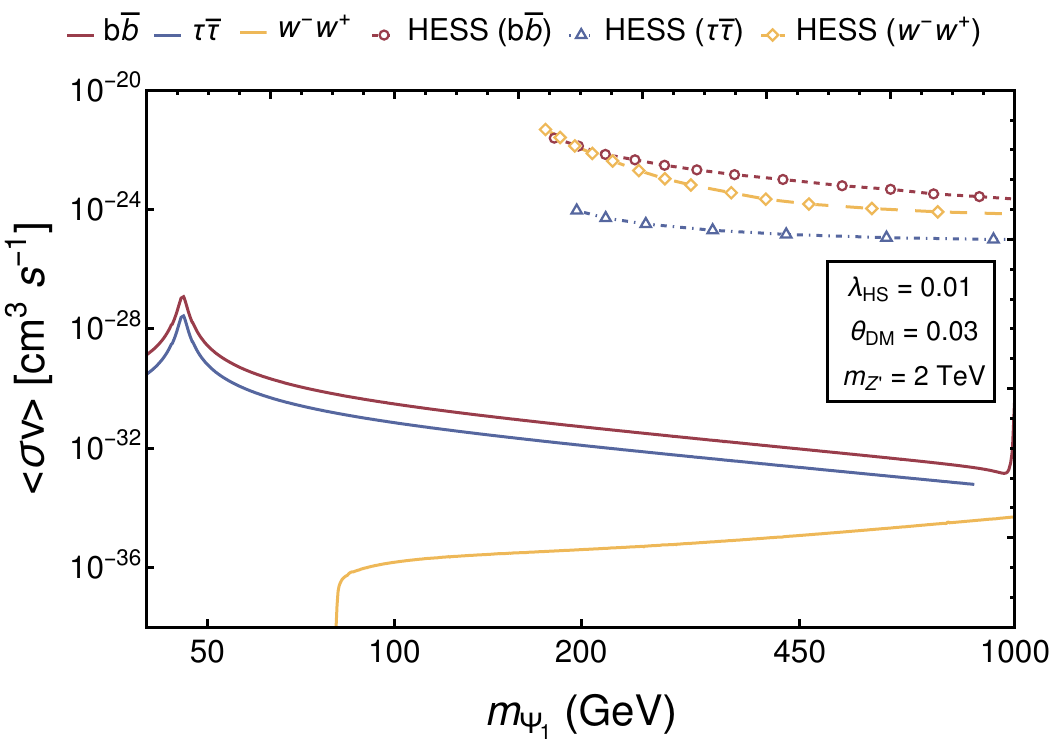}
\includegraphics[width=0.48 \textwidth]{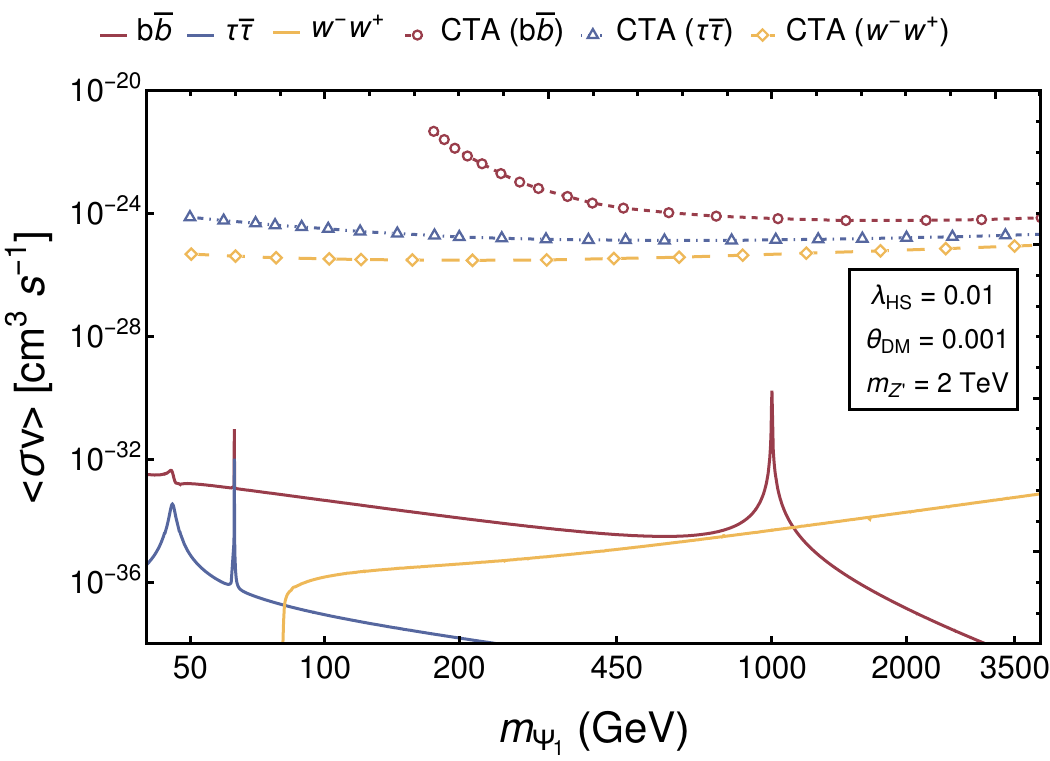}
\includegraphics[width=0.48 \textwidth]{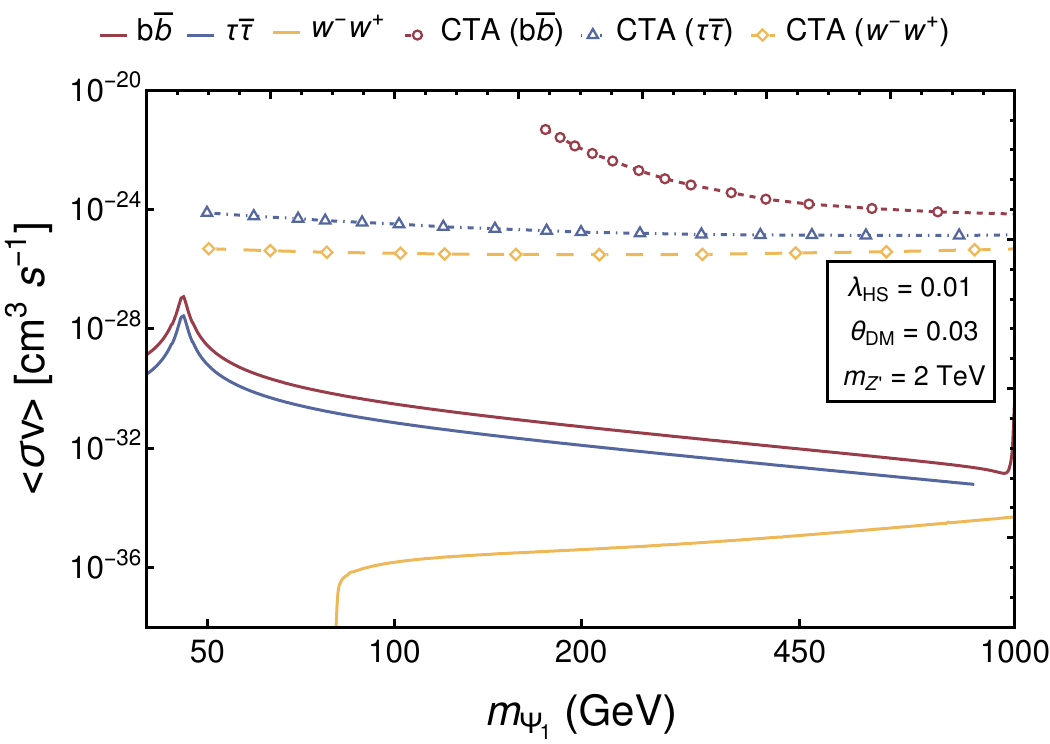}
\caption{Indirect detection constraints on the thermally averaged annihilation cross-section $ \langle \sigma v \rangle_{\rm ann} $ as a function of the dark matter mass $ m_{\Psi_1} $, for various annihilation channels. Current and projected upper limits from Fermi-LAT, H.E.S.S. and CTA observations based on diffuse gamma-ray measurements are also shown. The benchmark parameters used are $ \theta_{\rm DM} = 0.001 $ (\textbf{Left} column) and $\theta_{\rm DM}=0.03$ (\textbf{Right} column). Other parameters we consider are $ \lambda_{HS} = 0.01, \lambda_S=10^{-3}, m_{Z'}=2\text{ TeV},  y_\chi=0.01, y_\psi=0.05, y_\xi=0.1$.}
\label{fig:idd1}
\end{figure}

In Fig.~\ref{fig:idd1}, we show $\braket{\sigma v}_{\rm ann}$ for different final state annihilation channels, namely $ b\bar{b} $, $ \tau^+ \tau^- $ and $ W^+ W^- $, for various becnchmark paramater points. Current bounds from \texttt{Fermi-LAT}, \texttt{H.E.S.S} and future projection from \texttt{CTA} are shown. The \textbf{Left} and \textbf{Right} columns corresponds to $\theta_{\rm DM}=0.001$ and $0.03$ respectively. Other parameter values are mentioned in the caption. We see that the given benchmark point is not ruled out by indirect detection experiments. The $b\bar{b}$ annihilation cross section shows a narrow peak at the $Z'$ resonance, however, not reach the sensitivity of \texttt{H.E.S.S.} for the given parameter point. A thorough analysis of the annihilation cross-section near the $Z'$ resonance is required for a more definitive assessment of indirect constraints but is beyond the scope of this work. Overall, much of the DM parameter space in our model remains unconstrained by current indirect detection limits, though the $\Psi_1\Psi_1\rightarrow b\bar{b}$ channel may be further tested by future experiments.

\subsection{Collider bounds}
\label{sec:BLHC}
Collider experiments such as LEP and LHC provide stringent constraints on the parameter space of $U(1)_B$. In particular, the new gauge boson $Z'$ can be produced via quark-antiquark annihilation at colliders, which subsequently decays into a pair of leptons or jets. The strongest bounds arise from LHC searches for narrow resonances in the dilepton ($e^+e^-$, $\mu^+\mu^-$) and dijet final states. These analyses set lower limits on the $Z'$ mass $m_{Z'}$ as a function of the gauge coupling $g_B$. We utilize the results presented in~\cite{Taramati:2024kkn}, which are derived from the analyses performed in~\cite{ATLAS:2018qto, ATLAS:2020yat, ATLAS:2019lng, CMS:2018wxx}, valid in the mass range $m_{Z'} \in [500,1800]$ GeV. This study assumes $\Psi_1^+$ to be the heaviest among the four exotic fermionic states. Accordingly, we restrict our parameter space by adopting the hierarchy $y_\xi \gg y_\psi \gg y_\chi$. 
While broadly consistent, these constraints may differ slightly due to the inclusion of mixing between the singlet and doublet exotic fermions in our setup.

\begin{figure}[H]
\centering
\includegraphics[width=\textwidth]{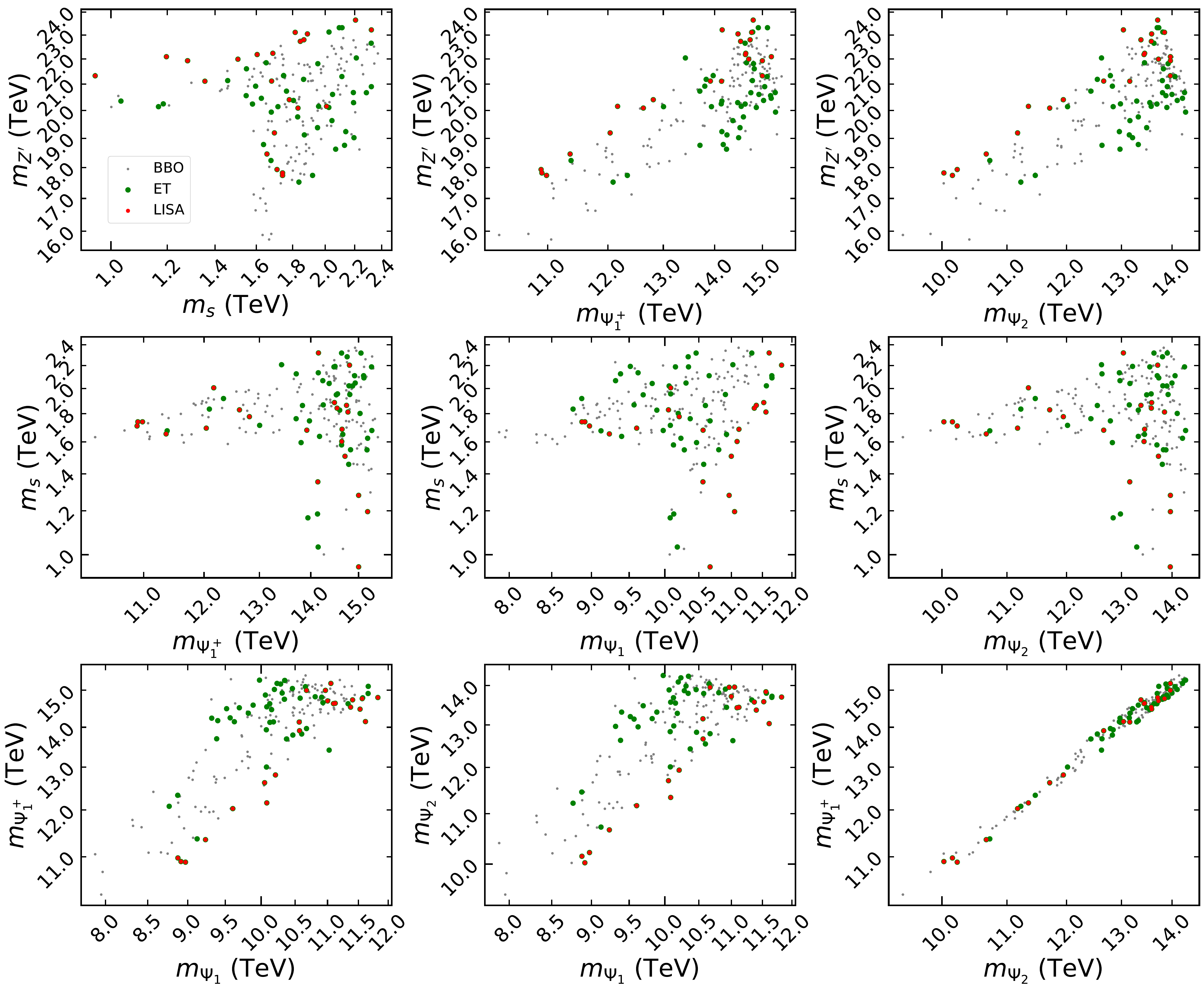}
\caption{ SNR $>$ 10 points with $\lambda_{HS}=0.01$ calculated for future GW experiments BBO, LISA and ET. Here, all the points satisfy the relic abundance bound of dark matter, constraints on oblique parameters from electroweak precision experiments within 1$\sigma$ range, current collider bounds and DM direct detection bounds. This plot shows the correlation among various physical masses of the model, accessible to future GW and laboratory experiments.}
\label{fig:SNR_stu_relic_lsh0_B1_-1_B2_2_thetaDM0.001_thetap0}
\end{figure}

In addition to the bounds on $m_{Z'}$ and $g_B$, the exotic fermion masses are also constrained from compressed MSSM search at ATLAS~\cite{Taramati:2024kkn, ATLAS:2019lng}. The ATLAS analysis sets a 95\% C.L. upper bound on the production cross-section of wino/bino-like chargino-neutralino pairs, which has been conservatively reinterpreted to constrain the $U(1)_B$ model by mapping the $\Psi_2^{\pm}\,\Psi_2$ production to the chargino-neutralino topology, projecting limits in the $m_{\Psi^\pm_2} = m_{\Psi_2}$ vs.\ $\Delta M(\Psi_2, \Psi_1)$ plane. The detailed analysis and plots of these bounds are given in~\cite{Taramati:2024kkn}. These bounds are taken into account while scanning the parameter space of the model in Fig.~\ref{fig:SNR_stu_relic_lsh0_B1_-1_B2_2_thetaDM0.001_thetap0} later.

In our model, the new charged fermions $\Psi^\pm$ and $\xi^\pm$ contribute to the $h \rightarrow \gamma\gamma$ decay amplitude at one loop. The total amplitude can be expressed as~\cite{Djouadi:2005gi,Spira:1995rr,Spira:2016ztx},

\begin{equation}
\mathcal{A}(h \to \gamma\gamma) = A_W + A_t + \sum_{F} A_F,
\end{equation}
where $A_W$ and $A_t$ denote the SM $W$-boson and top-quark loop contributions, respectively, and $A_F$ represents the contribution of a charged BSM fermion $F$. For a heavy fermion $F$ with mass $m_F \gg m_h$, the contribution simplifies to~\cite{Gunion:1989we,Djouadi:2005gi,Spira:1995rr},

\begin{equation}
A_F \simeq N_c Q_F^2 \frac{y_F v}{m_F} \cdot \frac{4}{3},
\end{equation}
where $N_c$ is the color factor, $Q_F$ is the electric charge of the fermion, $y_F$ is its Yukawa coupling to the Higgs, and $v = 246~\text{GeV}$ is the SM Higgs vacuum expectation value. Numerically, for the SM contributions, one has~\cite{Djouadi:2005gi,LHCHiggsCrossSectionWorkingGroup:2016ypw},

\begin{equation}
A_W \approx -8.3, \qquad A_t \approx +1.8, \qquad A_{\rm SM} = A_W + A_t \approx -6.5.
\end{equation}
The charged fermion masses are primarily generated from the $U(1)_B$ breaking scale $v_B$,

\begin{equation}
m_F \simeq y_F v_B,
\end{equation}
so that the ratio controlling the Higgs coupling to these fermions is

\begin{equation}
\frac{y_F v}{m_F} \simeq \frac{v}{v_B}.
\end{equation}
For the range of $v_B$ relevant for our analysis, $v_B \sim 10^4 - 10^5~\text{GeV}$, this ratio varies between

\begin{equation}
\frac{v}{v_B} \sim 2.5 \times 10^{-3} \text{ to } 2.5 \times 10^{-2},
\end{equation}
leading to BSM contributions to the amplitude

\begin{equation}
A_F \sim \mathcal{O}(10^{-2} - 10^{-1}),
\end{equation}
which remain much smaller than the SM amplitude $A_{\rm SM} \sim -6.5$. Consequently, the shift in the Higgs diphoton signal strength,

\begin{equation}
\mu_{\gamma\gamma} = \frac{|\mathcal{A}_{\rm SM} + \sum_F A_F|^2}{|A_{\rm SM}|^2} \simeq 1 + \mathcal{O}(10^{-3} - 10^{-2}),
\end{equation}
remains below the current experimental sensitivity. The latest combined ATLAS and CMS measurement gives~\cite{LHCHiggsCrossSectionWorkingGroup:2016ypw},
\begin{equation}
\mu_{\gamma\gamma} = 1.04 \pm 0.07,
\end{equation}
corresponding to a few-percent level precision. Even at the lower end $v_B \sim 10^4~\text{GeV}$, the BSM contributions are at most $\sim 1\%$, still below experimental sensitivity. This demonstrates that, in the parameter space considered, the new charged fermion contributions effectively decouple and do not impose any additional constraint on the Higgs diphoton decay. Therefore, we do not include the $h \rightarrow \gamma\gamma$ constraint explicitly in our numerical scans.

Future high-energy experiments will significantly enhance the sensitivity to this model across multiple channels. The HL-LHC will probe leptophobic $Z'$ bosons up to several TeV~\cite{Dobrescu:2021vak,Yu:2013wta,Ekstedt:2017tbo}, while searches for compressed electroweak fermions with soft leptons or disappearing tracks will constrain the $m_{\Psi_1}-\,\Delta M$ plane up to nearly 500 GeV~\cite{CMS-PAS-FTR-18-001}. FCC-hh will extend the $Z'$ mass reach to $\sim30$ TeV~\cite{Helsens:2642473} and lepton colliders like ILC will be sensitive to low-mass compressed states with high precision~\cite{Berggren:2021sns}. Long-lived particle (LLP) detectors like MATHUSLA~\cite{Curtin:2018mvb, Curtin:2023skh} and FASER~\cite{FASER:2018bac,FASER:2019aik,FASER:2019dxq} can also put constraints in the parameter space of the model. Finally, low-energy observables like parity violation (MOLLER, P2) and rare meson decays (NA62, KOTO) will test $m_{Z'}/g_B>10$ TeV~\cite{deMelo:2021ers,Dev:2021otb,Demiroglu:2024wys,Kitahara:2019lws}. These diverse and complementary probes make this an exceptionally testable and exciting model in upcoming experimental programs.

\section{Numerical analysis}\label{sec:numerical analysis}
In this section, we discuss the parameter scan of the model and the results obtained. We modified the packages \texttt{TransitionListener}\footnote{\href{https://github.com/tasicarl/TransitionListener}{https://github.com/tasicarl/TransitionListener}}~\cite{Ertas:2021xeh} and \texttt{CosmoTransitions}\footnote{\href{https://github.com/clwainwright/CosmoTransitions}{https://github.com/clwainwright/CosmoTransitions}} \cite{Wainwright:2011kj} to evaluate the phase transition parameters for different values of model parameters. We consider the free parameters in this model: $v_B, \lambda_S, g_B, y_\chi, y_\psi, y_\xi, \lambda_{HS}, \sin\theta_{\rm DM}, \sin\theta_p$ and derived parameters: $m_s, m_{Z'}, m_{\Psi_1}, m_{\Psi_2}, m_{\Psi_1^+}, m_{\Psi_2^+},\sin\theta, M_2, M_4$. The nature of the phase transition does not depend on the parameters $\sin\theta_{\rm DM}$ and $\sin\theta_p$; however, they affect DM phenomenology. First, we scan the following parameter space of our model for phase transition analysis,
\begin{align}
    \label{eq:params}
    v_B\in\left[10^3,10^6\right]\text{ GeV},\; \lambda_S\in\left[10^{-4},1\right],\;g_B\in\left[10^{-4},1\right],\;y_{\chi,\psi,\xi}\in\left[10^{-4},1\right]
\end{align}
along with the imposed condition: $y_\xi\gg y_\psi\gg y_\chi$ such that $\Psi_1$ is the DM candidate and $\Psi_1^+$ is heaviest, as discussed earlier. We fix the value for $\lambda_{HS}=0.01$ (see Appendix~\ref{app:beta} for details), which is related to scalar mixing $\sin\theta$. Additionally, as discussed in appendix~\ref{sec:relic}, we consider the resonant condition on the DM mass $m_{\Psi_1}\sim m_\chi=m_{Z'}/2$ in order to not overproduce dark matter, which is valid for small $\sin\theta_{\rm DM}$ values. We are interested in strong first-order phase transitions (SFOPT), which could give observable GW signals. We generate random parameter points within the given limits above and calculate different PT and GW parameters. We take $\sin\theta_p=0$ as discussed earlier.

We randomly generate around 20000 parameter points resulting in an FOPT. We make sure that each chosen point satisfies the conditions for the stability of the scalar potential given in Eq.~\eqref{eq:boundedness condition} and perturbativity limits given in Eq.~\eqref{eq:perturbativity condition}. Moreover, as described in Sec.~\ref {sec:effective potential}, we also make sure that the effective potential is bounded from below. Additionally, we also consider the running of the couplings in the model (see Appendix~\ref{app:beta} for RGEs) and accept parameter points only if they keep the Higgs potential stable and do not hit the Landau pole upto Planck scale. Out of the total number of random parameter points, we show 3 benchmark points (BP1, BP2, BP3, BP4) in Table~\ref{tab:BPFOPT} in terms of the physical masses. We show the example GW spectra for these benchmarks in Fig.~\ref{fig:BP_GW}, where we use the calculated phase transition parameters given in Table~\ref{tab:BPFOPTres}.


For DM phenomenology, we consider the same random points obtained in the above parameter scan and evaluate them considering two values for the DM mixing: $\sin\theta_{\rm DM}=0.001 \text{ and }0.03$. The charged sector mixing $\sin\theta_p$ is considered to be 0 as discussed earlier. With these values, we implement our model in \texttt{SARAH}\cite{Staub:2008uz}, \texttt{SPheno}~\cite{Porod:2011nf} and \texttt{micrOmegas}~\cite{Alguero:2023zol} to calculate the particle spectra and the DM relic abundance. For each of the parameter points, in addition to the DM relic density, we calculate $S, T, U$ parameters and $\sigma_{\rm SIDD}$ discussed in Sec .~\ref {sec:bounds}. We impose electroweak precision bounds on $S, T, U$ parameters within 1$\sigma$ confidence level, DM relic density bound $\Omega_{\Psi_1}h^2\lesssim 0.12$ and DM direct detection bounds from \texttt{LUX-ZEPLIN} (2025). We do not consider the indirect detection bounds, as they require more careful study. We also impose the current collider bounds on the parameter points coming from LHC described in Sec .~\ref {sec:bounds}. 

Finally, in Fig.~\ref{fig:SNR_stu_relic_lsh0_B1_-1_B2_2_thetaDM0.001_thetap0}, we show the viable parameter space of our model that satisfies all the theoretical and current experimental constraints, at the same time generating strong GW signals to be detected in future GW experiments such as LISA, ET and BBO. This parameter space will be sensitive to future GW, collider and DM experiments, demonstrating the complementarity of these experiments to probe this model. We showcase the viable parameter space as correlations among masses of various physical states in the model. The grey, green and red points are sensitive in BBO, ET and LISA, respectively, with SNR$>10$, implying $5\sigma$ C.L. of detection. For this figure, we consider $\lambda_{HS}=0.01,\, \sin\theta_{\rm DM}=0.001$. We find that taking $\sin\theta_{\rm DM}=0.03$ does not change the plots significantly.

\begin{figure}[H]
    \centering
    \includegraphics[width=0.49\linewidth]{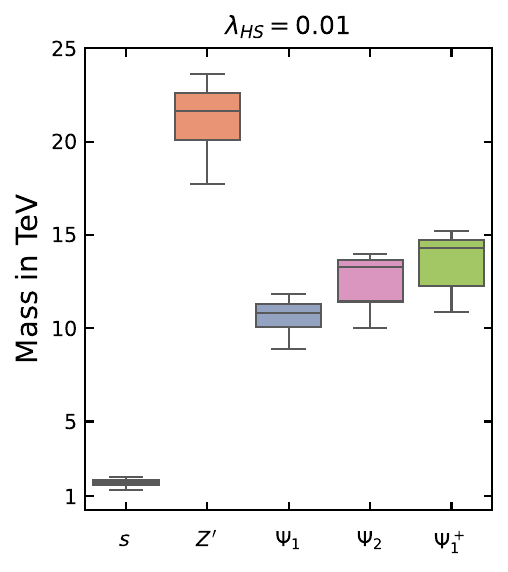}
    \caption{Box plots of physical masses for the parameter points shown in Fig.~\ref{fig:SNR_stu_relic_lsh0_B1_-1_B2_2_thetaDM0.001_thetap0}.
    The coloured boxes span the inter-quartile range (IQR) of the parameter points, the horizontal lines inside the boxes mark the median and the whiskers extend to the most extreme values within $1.5\times{\rm IQR}$. The diagrams summarise the typical mass scales realised by the viable parameter space of the model, which satisfies all the current theoretical and experimental constraints and can be probed in future collider studies as well as GW detectors.}
    \label{fig:box_plot}
\end{figure}

In Fig.~\ref{fig:box_plot}, we show box plots of the physical masses of the model. These plots summarize the distribution of viable physical masses within the model, displaying the median, interquartile range (IQR) and potential outliers. The box spans from the 25th to the 75th percentiles, with the central line indicating the median. Whiskers extend to data points within 1.5 times the IQR from the lower and upper quartiles. Points beyond this range are considered as outliers. These box plots are created considering the parameter points in Fig~\ref{fig:SNR_stu_relic_lsh0_B1_-1_B2_2_thetaDM0.001_thetap0} and they exhibit the statistics of the particle spectra satisfying all the current bounds discussed above and producing strong GW signals detectable at future GW observatories. Future colliders and DM detection experiments can shed light on the same parameter space, thereby complementing GW observations. In general, we find that DM mass $m_{\Psi_1}$ around 8 -- 12 TeV, $Z'$ mass around 16 -- 24 TeV and new scalar mass $m_s$ around 1 -- 2.5 TeV can be most interesting to test in the future GW, DM direct detection and collider experiments in synergy. Anything below these mass scales are ruled out primarily by LZ 2025 results while dark matter above $\sim 12$ TeV will not be detectable in LISA, BBO and ET.

\section{Conclusion and discussion}
\label{sec:conl}
We considered a minimal extension of the SM where baryon number is gauged with a $U(1)_B$ group. The anomaly cancellation is ensured by introducing new fermions with non-zero baryonic charges, such as $\chi$ (electric charge neutral and SM singlet), $\Psi$ ($SU(2)_L$ doublet and non-zero hypercharge), and $\xi$ (electrically charged and isospin singlet) to the particle content of the SM. These new fermions mix and give rise to four physical states $\Psi_1,\Psi_2,\Psi_1^+$ and $\Psi_2^+$, where the lightest $\Psi_1$ is found to be a stable dark matter candidate. Additionally, we add a complex scalar $S$ to break the $U(1)_B$ symmetry. 

We studied the dynamics of the phase transition in this prescribed framework. We considered the effective potential including one-loop Coleman-Weinberg corrections, counterterms to renormalize the potential, thermal corrections, and Daisy-resummation. Numerically, we examined various benchmark points leading to strong first-order phase transitions, which produce GWs through collisions and movements of nucleated bubbles in the thermal plasma of the early Universe. We have also checked the thermal relic abundance of dark matter and showed that in the resonance region $m_{\Psi_1}\sim m_{Z'}/2$, the DM satisfies the observed relic conditions while consistent with the prediction of gravitational wave signals. Additionally, we discussed constraints on this model from electroweak precision experiments, colliders, DM direct and indirect detection experiments. We also considered the running of the coupling constants of the model by solving the 1-loop renormalization group equations (RGEs) to find parameter values such that the Higgs potential remains stable and none of the couplings hit the Landau pole upto Planck scale. We scanned the parameter space of the model and found viable parameter points that satisfy all the current bounds while producing a strong enough GW signal to be detected in various future GW experiments such as LISA, ET, BBO, DECIGO, etc., with signal-to-noise ratio SNR $>10$. Future GW experiments and collider searches can simultaneously probe this viable parameter space and complement astrophysical experiments related to DM direct and indirect detection.  We showed the mass distributions of different physical states of our model as correlations in Fig.~\ref{fig:SNR_stu_relic_lsh0_B1_-1_B2_2_thetaDM0.001_thetap0} and as box plots in Fig.~\ref{fig:box_plot}. We showed that the parameter space of interest naturally favors masses of a few to tens of TeV for dark matter and the $Z'$ boson, while a few TeV masses for the scalar. They emerge as well-motivated targets for next-generation colliders operating at 100 TeV, where both the dark matter and $Z'$ resonances could be probed directly. These findings highlight the complementarity of GW, astrophysical, and collider frontiers in testing gauge-extended dark matter models.

\section*{Acknowledgement}
Authors thank Prof. Tao Han and Dr. Arnab Dasgupta for their valuable insights on various aspects of the paper. Taramati is supported by DST-INSPIRE (IF200289) fellowship. L.M. is supported by a UGC fellowship. Z.A.B. is supported by IoE fund at IIT Bombay.
\vspace{0.2cm}

\appendix
\section*{Appendix}
\section{Relic density of dark matter}
\label{sec:relic}
Here, we review the relic abundance of the dark matter candidate $\Psi_1$ in this model and explain the resonant condition that is considered throughout this paper. Mass hierarchies among $\Psi_1$, $\Psi_2$, $\Psi_{1}^+$ and $\Psi_{2}^+$ influence the relic abundance through self-annihilation and co-annihilation channels. Unlike in~\cite{Bringmann:2023iuz}, due to the presence of additional channels in our model, the DM mass need not exceed $m_s$ and $m_{Z'}$ to achieve the correct relic abundance. The cosmological evolution of exotic fermions follows the Boltzmann equation~\cite{Bhattacharya:2018fus},
\begin{equation}
\frac{dn}{dt} + 3 \mathcal{H} n = -\langle \sigma v\rangle_{\mathrm{eff}} (n^2 - n_{\mathrm{eq}}^2),
\end{equation}
where $ n = n_{\Psi_1}+n_{\Psi_2}+n_{\Psi^+_{1/2}}$. Since we consider $y_\chi\ll y_\psi\ll y_\xi$, the heavier particles $\Psi_2,\Psi^+_1,\Psi^+_2$ decays to the DM particle $\Psi_1$ such that eventualy $n=n_{\Psi_1}$.
The effective cross-section $ \langle \sigma v\rangle_{\mathrm{eff}} $ is~\cite{Taramati:2024kkn},
\begin{eqnarray}
    {\langle \sigma v\rangle}_{eff}&&= \frac{g_1^2}{g_{\rm eff}^2} {\langle \sigma v \rangle}_{\overline{\Psi_1}\Psi_1}+\frac{2 g_1 g_2}{g_{\rm eff}^2} {\langle \sigma v \rangle}_{\overline{\Psi_1}\Psi_2}\Big(1+\frac{\Delta M}{m_{\Psi_1}}\Big)^\frac{3}{2}  e^{-x \frac{\Delta M}{m_{\Psi_1}}} \nonumber \\
&& +\frac{2 g_1 g_3}{g_{\rm eff}^2} {\langle \sigma v \rangle}_{\overline{\Psi_1}\Psi_{i}^+}\Big(1+\frac{\Delta M}{m_{\Psi_1}}\Big)^\frac{3}{2} e^{-x \frac{\Delta M}{m_{\Psi_1}}} \nonumber \\
&& +\frac{2 g_2 g_3}{g_{\rm eff}^2} {\langle \sigma v \rangle}_{\overline{\Psi_2}\Psi_{i}^+}\Big(1+\frac{\Delta M}{m_{\Psi_1}}\Big)^3 e^{- 2 x \frac{\Delta M}{m_{\Psi_1}}} \nonumber \\
&& +\frac{g_2^2}{g_{\rm eff}^2} {\langle \sigma v \rangle}_{\overline{\Psi_2}\Psi_2}\Big(1+\frac{\Delta M}{m_{\Psi_1}}\Big)^3 e^{- 2 x \frac{\Delta M}{m_{\Psi_1}}} \nonumber \\
&& +\frac{g_3^2}{g_{eff}^2} {\langle \sigma v \rangle}_{{\Psi_{i}^+}\Psi_{i}^-}\Big(1+\frac{\Delta M}{m_{\Psi_1}}\Big)^3 e^{- 2 x \frac{\Delta M}{m_{\Psi_1}}},
\label{eq:vf-ann}
\end{eqnarray}
where $\Delta M=m_{\Psi_2}-m_{\Psi_1}$ is the mass difference between the DM and the next-to-lightest stable particle and $ g_{\rm eff} $ is,
\begin{equation}
g_{\rm eff} = g_1 + g_2 \left(1+\frac{\Delta M}{m_{\Psi_1}}\right)^{3/2} e^{-x \frac{\Delta M}{m_{\Psi_1}}} + g_3 \left(1+\frac{\Delta M}{m_{\Psi_1}}\right)^{3/2} e^{-x \frac{\Delta M}{m_{\Psi_1}}},
\end{equation}
where $g_1,g_2,g_3$ are the degrees of fredom of $\Psi_1,\Psi_2,\Psi^+_{i}$ (where $i=1,2$). The final relic abundance \cite{Griest:1990kh} is,
\begin{equation}
\Omega_{\Psi_1}h^2 = \frac{1.09 \times 10^9 \text{GeV}^{-1}}{g_*^{1/2}M_{\rm Pl}}\frac{1}{J(x_f)},
\end{equation}
with $ J(x_f) $ defined as:
\begin{equation}
J(x_f) = \int_{x_f}^{\infty} \frac{\langle \sigma |v| \rangle_{\mathrm{eff}}}{x^2} dx.
\end{equation}

It is seen that the DM relic abundance is usually above the experimental limit $0.12$ for most of the parameter space except at the \textit{resonance} regimes where the mass of DM is approximately half of the mass of the mediator of the annihilation channel. This enhances the annihilation cross section of DM via on-shell production of the mediator. We find in this model that such resonance in the annihilation via $Z'$ mediator allows the DM relic to be within the experimental limit~\cite{FileviezPerez:2019jju, Duerr:2016tmh}. Therefore, in our analysis, we consider the resonant condition $m_{\Psi_1}\sim m_{Z'}/2$ such that the DM does not overclose the Universe.
\section{One-Loop Beta Functions}
\label{app:beta}
The renormalisation group equations (RGEs) of a parameter $p_i$ can be written as
\begin{equation}
    \frac{dp_i}{dt}=\frac{1}{16\pi^2} \beta_{p_i}
\end{equation}
where $\beta_{p_i}$ is the corresponding beta function and $t=\ln{\left ( \frac{\mu}{\rm GeV} \right )}$ is the variable related to the renormalisation scale $\mu$. We define the threshold step functions for the BSM fields
\begin{equation}
\Theta_\Psi \equiv \theta(\mu - m_\Psi),\qquad
\Theta_\xi \equiv \theta(\mu - m_\xi),\qquad
\Theta_\chi \equiv \theta(\mu - m_\chi).
\end{equation}
For compactness we use the conservative shorthand
\begin{equation}
\Theta_B^{\rm sum} \equiv \Theta_\Psi+\Theta_\xi
\end{equation}
where the + operator signifies the \textit{OR} operation. The above factor is multiplied to the \(U(1)_B\) (and mixed \(U(1)\)–\(U(1)_B\)) pure-gauge contributions to switch them on only when the baryon-charged species are active. The one-loop beta functions for various couplings of the model are given below.

\medskip
\subsection{Gauge couplings}
\begin{align}
\beta_{g_1}^{(1)} &= \frac{1}{10}\Big(16\sqrt{10}\,g_{1}^{2} g_{BY} + 53 g_{1}^{3} 
+ g_1\big(285 g_{BY}^{2} + 53 g_{YB}^{2} + 8\sqrt{10}\, g_{YB} g_B\big)\,(\Theta_B^{\rm sum}) \nonumber\\
&\qquad\qquad + g_{YB}\big(285 g_B + 8\sqrt{10}\,g_{YB}\big)g_{BY}\,(\Theta_B^{\rm sum})\Big),
\\[6pt]
\beta_{g_{YB}}^{(1)} &= \frac{1}{10}\Big(g_{1}^{2}\big(53 g_{YB} + 8\sqrt{10}\, g_B\big)\,(\Theta_\Psi+\Theta_\xi) 
+ g_1\big(285 g_B + 8\sqrt{10}\, g_{YB}\big)g_{BY}\,(\Theta_B^{\rm sum}) \nonumber\\
&\qquad\qquad + g_{YB}\big(16\sqrt{10}\, g_{YB} g_B + 285 g_{B}^{2} + 53 g_{YB}^{2}\big)\,(\Theta_B^{\rm sum})\Big),
\\[6pt]
\beta_{g_2}^{(1)} &= -\tfrac{5}{2} g_{2}^{3},
\\[6pt]
\beta_{g_3}^{(1)} &= -7 g_{3}^{3},
\\[6pt]
\beta_{g_B}^{(1)} &= \frac{1}{10}\Big(16\sqrt{10}\, g_{YB} g_{B}^{2} + 285 g_{B}^{3} + 53 g_{YB}^{2} g_B 
+ g_{YB} g_{BY}\big(53 g_1 + 8\sqrt{10}\, g_{BY}\big) \nonumber\\
&\qquad\qquad + g_B g_{BY}\big(285 g_{BY} + 8\sqrt{10}\, g_1\big)\Big)\,(\Theta_B^{\rm sum}),
\\[6pt]
\beta_{g_{BY}}^{(1)} &= \frac{1}{10}\Big(285 g_{BY}\big(g_{B}^{2} + g_{BY}^{2}\big) + 53 g_{1}^{2} g_{BY} + 53 g_1 g_{YB} g_B \nonumber\\
&\qquad\qquad + 8\sqrt{10}\, g_1\big(2 g_{BY}^{2} + g_{B}^{2}\big) + 8\sqrt{10}\, g_{YB} g_B g_{BY}\Big)\,(\Theta_B^{\rm sum}).
\end{align}

\subsection{Scalar quartic couplings}
\begin{align}
\beta_{\lambda_H}^{(1)} &= \frac{27}{200} g_1^4 + \frac{9}{20} g_1^2 g_2^2 + \frac{9}{8} g_2^4 
-\frac{9}{5} g_1^2 \lambda_H -\frac{9}{5} g_{YB}^2 \lambda_H -9 g_2^2 \lambda_H \nonumber\\
&\quad +24\lambda_H^2 + \lambda_{HS}^2 
+ 4\lambda_H |y_1|^2 \,\Theta_\Psi + 4\lambda_H |y_2|^2 \,\Theta_\Psi + 4\lambda_H |y_3|^2 \,\Theta_\Psi + 4\lambda_H |y_4|^2 \,\Theta_\Psi \nonumber\\
&\quad -2|y_1|^4 \,\Theta_\Psi -2|y_2|^4 \,\Theta_\Psi -2|y_3|^4 \,\Theta_\Psi -2|y_4|^4 \,\Theta_\Psi \nonumber\\
&\quad +12\lambda_H \mathrm{Tr}(Y_d Y_d^\dagger) +4\lambda_H \mathrm{Tr}(Y_e Y_e^\dagger) +12\lambda_H \mathrm{Tr}(Y_u Y_u^\dagger) \nonumber\\
&\quad -6\mathrm{Tr}(Y_d Y_d^\dagger Y_d Y_d^\dagger) -2\mathrm{Tr}(Y_e Y_e^\dagger Y_e Y_e^\dagger) -6\mathrm{Tr}(Y_u Y_u^\dagger Y_u Y_u^\dagger),
\\[8pt]
\beta_{\lambda_{HS}}^{(1)} &= -\frac{9}{10} g_1^2 \lambda_{HS} -\frac{9}{10} g_{YB}^2 \lambda_{HS} -\frac{9}{2} g_2^2 \lambda_{HS} 
-81 g_B^2 \lambda_{HS}\,(\Theta_B^{\rm sum}) -81 g_{BY}^2 \lambda_{HS}\,(\Theta_B^{\rm sum}) \nonumber\\
&\quad +12\lambda_H\lambda_{HS} +8\lambda_S\lambda_{HS} +4\lambda_{HS}^2 
+2\lambda_{HS} |y_2|^2 \,\Theta_\Psi +2\lambda_{HS} |y_3|^2 \,\Theta_\Psi +2\lambda_{HS} |y_4|^2 \,\Theta_\Psi \nonumber\\
&\quad +4\lambda_{HS} |y_{\Psi}|^2 \,\Theta_\Psi +2\lambda_{HS} |y_{\xi}|^2 \,\Theta_\xi \nonumber\\
&\quad +2 y_1^*\Big(-2 y_{\Psi} y_{\xi} y_2^* + y_1\big(-2|y_{\Psi}|^2 -2|y_{\xi}|^2 + \lambda_{HS}\big)\Big)\,\Theta_\Psi \nonumber\\
&\quad -4 y_{\chi} y_{\Psi} y_3^* y_4^* \,(\Theta_\chi\Theta_\Psi) 
-4 y_{\Psi}|y_2|^2 y_{\Psi}^*\,\Theta_\Psi -4 y_{\Psi}|y_3|^2 y_{\Psi}^*\,\Theta_\Psi -4 y_{\Psi}|y_4|^2 y_{\Psi}^*\,\Theta_\Psi \nonumber\\
&\quad +2 y_{\chi}^*\Big(-2 y_3 y_4 y_{\Psi}^* -2 y_{\chi} |y_3|^2 -2 y_{\chi}|y_4|^2 + \lambda_{HS} y_{\chi}\Big)\,\Theta_\chi \nonumber\\
&\quad -4 y_{\xi} |y_2|^2 y_{\xi}^*\,\Theta_\xi -4 y_1 y_2 y_{\Psi}^* y_{\xi}^*\,(\Theta_\Psi\Theta_\xi) \nonumber\\
&\quad +6\lambda_{HS} \mathrm{Tr}(Y_d Y_d^\dagger) +2\lambda_{HS} \mathrm{Tr}(Y_e Y_e^\dagger) +6\lambda_{HS} \mathrm{Tr}(Y_u Y_u^\dagger),
\\[8pt]
\beta_{\lambda_S}^{(1)} &= \frac{2187}{2} g_B^4 \,(\Theta_B^{\rm sum}) -162 g_B^2 \lambda_S \,(\Theta_B^{\rm sum}) -162 g_{BY}^2 \lambda_S \,(\Theta_B^{\rm sum}) \nonumber\\
&\quad +20\lambda_S^2 +2\lambda_{HS}^2 
+4\lambda_S |y_{\chi}|^2 \,\Theta_\chi +8\lambda_S |y_{\Psi}|^2 \,\Theta_\Psi +4\lambda_S |y_{\xi}|^2 \,\Theta_\xi \nonumber\\
&\quad -2|y_{\chi}|^4 \,\Theta_\chi -4|y_{\Psi}|^4 \,\Theta_\Psi -2|y_{\xi}|^4 \,\Theta_\xi.
\end{align}

\subsection{Yukawa couplings}
For each Yukawa RGE the extra fermion dependent pieces are multiplied by the threshold(s) of the fermion(s) appearing in that term; the $U(1)_B$-gauge pieces are multiplied by the conservative $\Theta_B^{\rm sum}$ where indicated.

\begin{align}
\beta_{Y_u}^{(1)} &= -\tfrac{3}{2}\big(-Y_u Y_u^\dagger Y_u + Y_u Y_d^\dagger Y_d\big) \nonumber\\
&\quad + Y_u\Big(-\tfrac{17}{20} g_1^2 -\tfrac{17}{20} g_{YB}^2 -\tfrac{9}{4}g_2^2 -8 g_3^2 -\sqrt{\tfrac{5}{2}} g_{YB} g_B - g_B^2 \nonumber\\
&\qquad\qquad -\sqrt{\tfrac{5}{2}} g_1 g_{BY} - g_{BY}^2 + |y_1|^2\Theta_\Psi + |y_2|^2\Theta_\Psi + |y_3|^2\Theta_\Psi + |y_4|^2\Theta_\Psi \nonumber\\
&\qquad\qquad + 3\mathrm{Tr}(Y_d Y_d^\dagger) + \mathrm{Tr}(Y_e Y_e^\dagger) + 3\mathrm{Tr}(Y_u Y_u^\dagger)\Big),
\\[6pt]
\beta_{Y_d}^{(1)} &= \tfrac{3}{2}\big(-Y_d Y_u^\dagger Y_u + Y_d Y_d^\dagger Y_d\big) \nonumber\\
&\quad + Y_d\Big(-\tfrac{1}{4} g_1^2 -\tfrac{1}{4} g_{YB}^2 -\tfrac{9}{4} g_2^2 -8 g_3^2 + \tfrac{1}{\sqrt{10}} g_{YB} g_B - g_B^2 \nonumber\\
&\qquad\qquad + \tfrac{1}{\sqrt{10}} g_1 g_{BY} - g_{BY}^2 + |y_1|^2\Theta_\Psi + |y_2|^2\Theta_\Psi + |y_3|^2\Theta_\Psi + |y_4|^2\Theta_\Psi \nonumber\\
&\qquad\qquad + 3\mathrm{Tr}(Y_d Y_d^\dagger) + \mathrm{Tr}(Y_e Y_e^\dagger) + 3\mathrm{Tr}(Y_u Y_u^\dagger)\Big),
\\[6pt]
\beta_{Y_e}^{(1)} &= \tfrac{3}{2} Y_e Y_e^\dagger Y_e + Y_e\Big(3\mathrm{Tr}(Y_d Y_d^\dagger) + 3\mathrm{Tr}(Y_u Y_u^\dagger) -\tfrac{9}{4} g_1^2 -\tfrac{9}{4} g_2^2 -\tfrac{9}{4} g_{YB}^2 \nonumber\\
&\qquad\qquad + |y_1|^2\Theta_\Psi + |y_2|^2\Theta_\Psi + |y_3|^2\Theta_\Psi + |y_4|^2\Theta_\Psi + \mathrm{Tr}(Y_e Y_e^\dagger)\Big).
\end{align}

\begin{align}
\beta_{y_4}^{(1)} &= \frac{1}{20} y_4\Big(-9 g_1^2 -9 g_{YB}^2 -45 g_2^2 
-36\sqrt{10}\, g_{YB} g_B\,(\Theta_B^{\rm sum}) -720 g_B^2\,(\Theta_B^{\rm sum}) \nonumber\\
&\quad -36\sqrt{10}\, g_1 g_{BY}\,(\Theta_B^{\rm sum}) -720 g_{BY}^2\,(\Theta_B^{\rm sum})
+10|y_\chi|^2\Theta_\chi +20|y_1|^2\Theta_\Psi \nonumber\\
&\quad -10|y_2|^2\Theta_\Psi +20|y_3|^2\Theta_\Psi +50|y_4|^2\Theta_\Psi +10|y_\Psi|^2\Theta_\Psi \nonumber\\
&\quad +60\mathrm{Tr}(Y_d Y_d^\dagger) +20\mathrm{Tr}(Y_e Y_e^\dagger) +60\mathrm{Tr}(Y_u Y_u^\dagger)\Big),
\\[8pt]
\beta_{y_2}^{(1)} &= \frac{1}{20} y_2\Big(-45 g_1^2 -45 g_{YB}^2 -45 g_2^2 
-108\sqrt{10}\, g_{YB} g_B\,(\Theta_B^{\rm sum}) -720 g_B^2\,(\Theta_B^{\rm sum}) \nonumber\\
&\quad -108\sqrt{10}\, g_1 g_{BY}\,(\Theta_B^{\rm sum}) -720 g_{BY}^2\,(\Theta_B^{\rm sum})
+20|y_1|^2\Theta_\Psi +50|y_2|^2\Theta_\Psi \nonumber\\
&\quad +20|y_3|^2\Theta_\Psi -10|y_4|^2\Theta_\Psi +10|y_\Psi|^2\Theta_\Psi +10|y_\xi|^2\Theta_\xi \nonumber\\
&\quad +60\mathrm{Tr}(Y_d Y_d^\dagger) +20\mathrm{Tr}(Y_e Y_e^\dagger) +60\mathrm{Tr}(Y_u Y_u^\dagger)\Big),
\\[8pt]
\beta_{y_\chi}^{(1)} &= \tfrac{1}{2} y_\chi\Big(2|y_3|^2\Theta_\Psi + 2|y_4|^2\Theta_\Psi + 2|y_\xi|^2\Theta_\xi 
-45 g_B^2\,(\Theta_B^{\rm sum}) -45 g_{BY}^2\,(\Theta_B^{\rm sum}) \nonumber\\
&\qquad\qquad +4|y_\chi|^2\Theta_\chi +4|y_\Psi|^2\Theta_\Psi\Big),
\\[8pt]
\beta_{y_\xi}^{(1)} &= \tfrac{1}{10} y_\xi\Big(-36 g_1^2 -36 g_{YB}^2 -18\sqrt{10}\, g_{YB} g_B\,(\Theta_B^{\rm sum}) -225 g_B^2\,(\Theta_B^{\rm sum}) \nonumber\\
&\quad -18\sqrt{10}\, g_1 g_{BY}\,(\Theta_B^{\rm sum}) -225 g_{BY}^2\,(\Theta_B^{\rm sum})
+10|y_\chi|^2\Theta_\chi +10|y_1|^2\Theta_\Psi \nonumber\\
&\quad +10|y_2|^2\Theta_\Psi +20|y_\Psi|^2\Theta_\Psi +20|y_\xi|^2\Theta_\xi\Big),
\\[8pt]
\beta_{y_\Psi}^{(1)} &= \tfrac{1}{10} y_\Psi\Big(-9 g_1^2 -9 g_{YB}^2 -45 g_2^2 
-9\sqrt{10}\, g_{YB} g_B\,(\Theta_B^{\rm sum}) -225 g_B^2\,(\Theta_B^{\rm sum}) \nonumber\\
&\quad -9\sqrt{10}\, g_1 g_{BY}\,(\Theta_B^{\rm sum}) -225 g_{BY}^2\,(\Theta_B^{\rm sum})
+10|y_\chi|^2\Theta_\chi +5|y_1|^2\Theta_\Psi \nonumber\\
&\quad +5|y_2|^2\Theta_\Psi +5|y_3|^2\Theta_\Psi +5|y_4|^2\Theta_\Psi +30|y_\Psi|^2\Theta_\Psi +10|y_\xi|^2\Theta_\xi\Big),
\\[8pt]
\beta_{y_3}^{(1)} &= \frac{1}{20} y_3\Big(-9 g_1^2 -9 g_{YB}^2 -45 g_2^2 +18\sqrt{10}\, g_{YB} g_B\,(\Theta_B^{\rm sum}) -180 g_B^2\,(\Theta_B^{\rm sum}) \nonumber\\
&\quad +18\sqrt{10}\, g_1 g_{BY}\,(\Theta_B^{\rm sum}) -180 g_{BY}^2\,(\Theta_B^{\rm sum})
+10|y_\chi|^2\Theta_\chi -10|y_1|^2\Theta_\Psi \nonumber\\
&\quad +20|y_2|^2\Theta_\Psi +50|y_3|^2\Theta_\Psi +20|y_4|^2\Theta_\Psi +10|y_\Psi|^2\Theta_\Psi \nonumber\\
&\quad +60\mathrm{Tr}(Y_d Y_d^\dagger) +20\mathrm{Tr}(Y_e Y_e^\dagger) +60\mathrm{Tr}(Y_u Y_u^\dagger)\Big),
\\[8pt]
\beta_{y_1}^{(1)} &= \frac{1}{20} y_1\Big(-45 g_1^2 -45 g_{YB}^2 -45 g_2^2 +54\sqrt{10}\, g_{YB} g_B\,(\Theta_B^{\rm sum}) -180 g_B^2\,(\Theta_B^{\rm sum}) \nonumber\\
&\quad +54\sqrt{10}\, g_1 g_{BY}\,(\Theta_B^{\rm sum}) -180 g_{BY}^2\,(\Theta_B^{\rm sum})
+50|y_1|^2\Theta_\Psi +20|y_2|^2\Theta_\Psi \nonumber\\
&\quad -10|y_3|^2\Theta_\Psi +20|y_4|^2\Theta_\Psi +10|y_\Psi|^2\Theta_\Psi +10|y_\xi|^2\Theta_\xi \nonumber\\
&\quad +60\mathrm{Tr}(Y_d Y_d^\dagger) +20\mathrm{Tr}(Y_e Y_e^\dagger) +60\mathrm{Tr}(Y_u Y_u^\dagger)\Big).
\end{align}

\subsection{Mass terms}
\begin{align}
\beta_{\mu_H^2}^{(1)} &= 2\lambda_{HS}\,\mu_S^2\,\Theta_S -\frac{9}{10} g_1^2 \mu_H^2 -\frac{9}{10} g_{YB}^2 \mu_H^2 -\frac{9}{2} g_2^2 \mu_H^2 +12\lambda_H \mu_H^2 \nonumber\\
&\quad +2\mu_H^2 |y_1|^2\Theta_\Psi +2\mu_H^2 |y_2|^2\Theta_\Psi +2\mu_H^2 |y_3|^2\Theta_\Psi +2\mu_H^2 |y_4|^2\Theta_\Psi \nonumber\\
&\quad +6\mu_H^2 \mathrm{Tr}(Y_d Y_d^\dagger) +2\mu_H^2 \mathrm{Tr}(Y_e Y_e^\dagger) +6\mu_H^2 \mathrm{Tr}(Y_u Y_u^\dagger),
\\[8pt]
\beta_{\mu_S^2}^{(1)} &= 2\mu_S^2 |y_\chi|^2\Theta_\chi + 2\mu_S^2 |y_\xi|^2\Theta_\xi +4\lambda_{HS}\mu_H^2 +4\mu_S^2 |y_\Psi|^2\Theta_\Psi \nonumber\\
&\quad -81 g_{BY}^2 \mu_S^2 \,(\Theta_B^{\rm sum}) -81 g_B^2 \mu_S^2\,(\Theta_B^{\rm sum}) +8\lambda_S \mu_S^2.
\end{align}

\begin{figure}[H]
    \centering
    \includegraphics[width=0.49\linewidth]{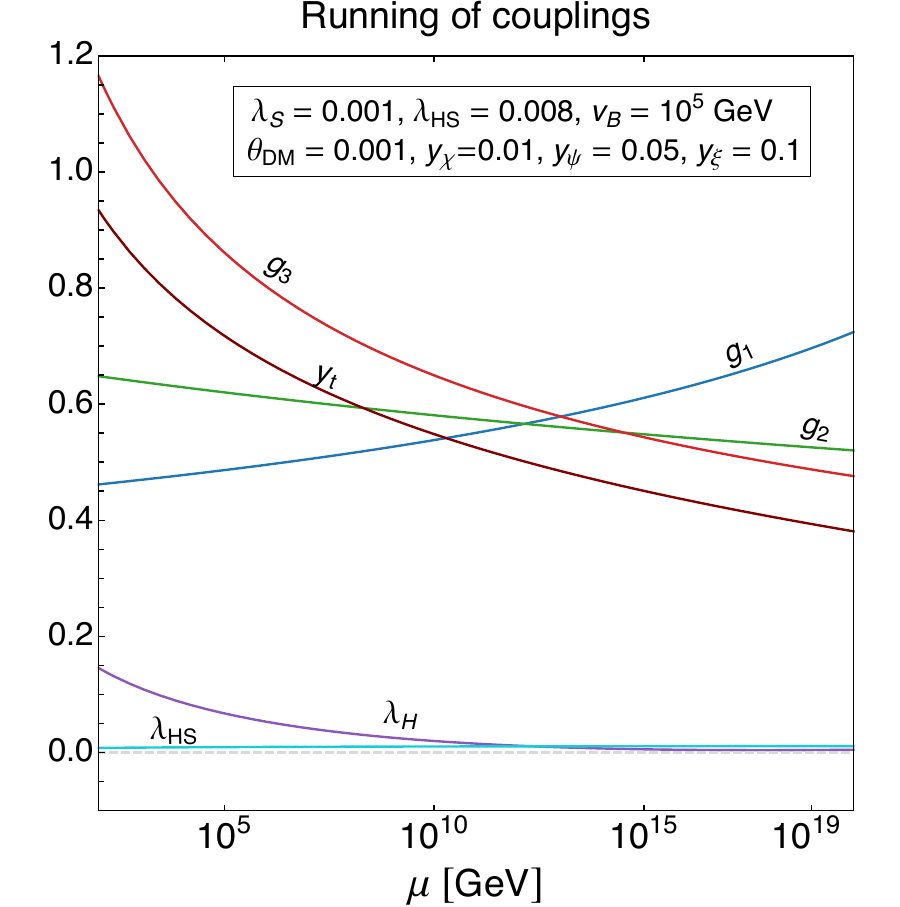}
    \includegraphics[width=0.49\linewidth]{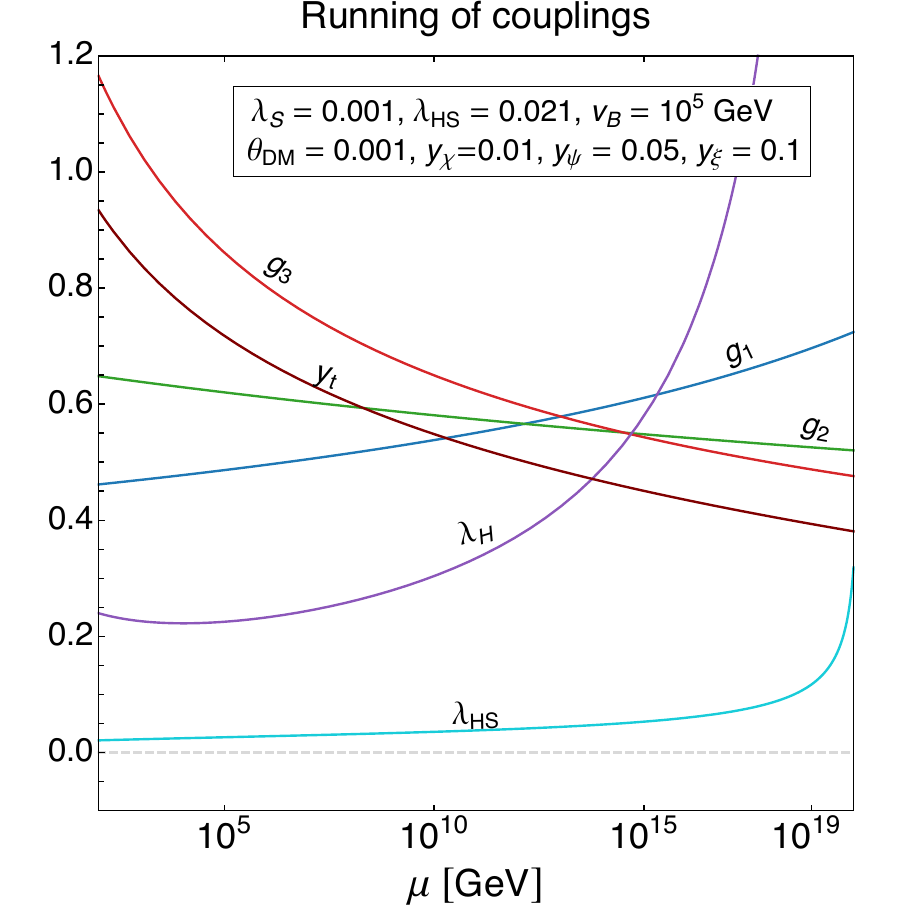}
    \caption{Running of the coupling constants in our model for representative parameter values. The left plot shows the minimum $\lambda_{HS}$ allowed such that the Higgs potential remains stable upto Planck scale, meaning $\lambda_H$ remains positive. The plot on the right shows the maximum value of $\lambda_{HS}$ such that $\lambda_H$ does not hit the Laundau pole breaking the perturbativity of the model.}
    \label{fig:rge}
\end{figure}

\subsection{Running of the couplings}
 \par For the numerical evaluation of the renormalization group, we adopt a simplified form of all the beta functions. In this approach, all SM couplings are neglected except for the gauge couplings, the top-quark Yukawa coupling, and the Higgs quartic coupling. In Fig.~\ref{fig:rge}, we show the running of the couplings from electroweak scale upto Planck scale for representative parameter values $\lambda_S=10^{-3}$, $\theta_{\rm DM}=0.001$, $v_B=10^5\text{ GeV}$, $y_\chi=0.01$, $y_\psi=0.05$, $y_\xi=0.1$ while varying the portal coupling $\lambda_{HS}$. All other dimensionless couplings other than those shown in the figure remain almost constant upto Planck scale. In Fig.~\ref{fig:rge}, the plot on the left shows the minimum value of $\lambda_{HS}\sim0.008$ for which $\lambda_H$ remains positive throughout, which means that the Higgs potential is stable. The plot on the right demonstrate the maximum value of $\lambda_{HS}=0.021$ allowed beyond which the perturbativity limit of the model is broken since $\lambda_H$ hits the Laundau pole. This allowed range of $\lambda_{HS}$ changes depending on values of $\lambda_S$ and $\theta_{\rm DM}$. From a phenomenological perspective, we are interested in $\lambda_S$ and $\theta_{\rm DM}$ values close to $10^{-3}$ and $0.001$ respectively, hence for the analysis of this paper we choose $\lambda_{HS}=0.01$. For the random parameter scan (see Fig.~\ref{fig:SNR_stu_relic_lsh0_B1_-1_B2_2_thetaDM0.001_thetap0}), we impose the stability of the Higgs potential and perturbativity conditions from RGEs as described in the main text. 
 
We highlight that imposing vacuum stability and perturbativity up to the Planck scale is a conservative requirement to ensure UV robustness. This assumption is not strictly necessary, and the effective theory may remain valid only up to a lower cutoff scale set by the heaviest states in the spectrum, which in our model is around the $U(1)_B$ breaking scale $v_B$. Relaxing this requirement would enlarge the allowed parameter space, e.g. a larger value of $\lambda_{HS}$, however, without qualitatively affecting our conclusions regarding the phase transition dynamics and the resulting gravitational wave signals.


\newpage

\bibliographystyle{apsrev4-2}
\bibliography{ref-merge}

\end{document}